\documentclass[final,5p,times,twocolumn]{elsarticle}

\usepackage[utf8x]{inputenc}
\usepackage[super]{nth}
\usepackage{amsmath, amsfonts}
\usepackage{enumitem}
\usepackage{dsfont}
\usepackage[absolute,overlay]{textpos}

\usepackage[hyphens]{url}
\usepackage{hyperref}
\usepackage{makecell}
\usepackage{multirow}
\usepackage{booktabs} 
\usepackage{graphicx}
\usepackage{subcaption}
\usepackage{comment}
\usepackage{xspace}
\usepackage{color}
\usepackage{flushend}

\usepackage{arydshln}
\usepackage{amsmath}
\usepackage{xcolor}

\usepackage{color, colortbl}

\newcommand{\whiteColor}[1]{{\color{white}[CH: #1]}}

\makeatletter
\def\ps@pprintTitle{%
  \let\@oddhead\@empty
  \let\@evenhead\@empty
  \let\@oddfoot\@empty
  \let\@evenfoot\@oddfoot
}
\makeatother

\begin{document}

\title{On the Dynamics of Political Discussions on Instagram: A Network Perspective}

\author[1,2,3]{Carlos H. G. Ferreira}
\author[2]{Fabricio Murai}
\author[2]{Ana P. C. Silva}
\author[2]{Jussara M. Almeida}
\author[3]{\\Martino Trevisan}
\author[3]{Luca Vassio}
\author[3]{Marco Mellia}
\author[4]{Idilio Drago}

\address[1]{Universidade Federal de Ouro Preto, Brazil}
\address[2]{Universidade Federal de Minas Gerais, Brazil}
\address[3]{Politecnico di Torino, Italy}
\address[4]{University of Turin, Italy}

\begin{abstract}
Instagram  has been increasingly used as a source of information especially among the youth. As a result, political figures now leverage the platform to spread opinions and political agenda. We here analyze online discussions   on Instagram, notably in political topics, from a network perspective. Specifically, we investigate  the emergence of  communities of co-commenters, that is, groups of users who often interact by commenting on the same posts and may be driving the ongoing online discussions. In particular, we are interested in \emph{salient co-interactions}, i.e., interactions of co-commenters that occur more often than expected by chance and under independent behavior. Unlike casual and accidental co-interactions which normally happen in large volumes, salient co-interactions are key elements driving the online discussions and, ultimately, the information dissemination. We base our study on the analysis of 10 weeks of data centered around major elections in Brazil and Italy, following both politicians and other celebrities. We extract and characterize the communities of co-commenters in terms of topological structure, properties of the discussions carried out by community members, and how some community properties, notably community membership and topics, evolve over time. We show that communities discussing political topics tend to be more engaged in the debate by writing longer comments, using more emojis, hashtags and negative words than in other subjects. Also, communities built around political discussions tend to be more dynamic, although top commenters remain active and preserve community membership over time.  Moreover, we  observe  a great diversity in discussed topics over time: whereas some topics attract attention only momentarily, others, centered around more fundamental political discussions, remain consistently active over time.

%Moreover,  salient co-interactions often arise around provoking posts,
%but some co-interactions centered around fundamental political discussions last longer. 

%\ju{ These findings are not 100\% aligned with the intro. Check what is the correct message and revise.} \ch{Poderia ser: We show that in politics, multiple communities engage with the same influencer (or a subset of them) but are typically driven by different topics. Also,  communities discussing political topics tend to be structurally stronger and use longer comments, more emojis, hashtags and negative words than in other subjects. Last, communities built around political discussions tend to be more dynamic, although top commenters tend to remain active and preserve community membership over time.}
\end{abstract}


\maketitle

\TPshowboxestrue
\TPMargin{0.3cm}
\begin{textblock*}{18cm}(1.5cm,0.8cm)
\small
\bf
\definecolor{myRed}{rgb}{0.55,0,0}
\color{myRed}
\noindent
Please cite this article as: Carlos H. G. Ferreira, Fabricio Murai, Ana P. C. Silva, Jussara M. Almeida, Martino Trevisan, Luca Vassio, Marco Mellia, Idilio Drago. On the Dynamics of Political Discussions on Instagram: A Network Perspective. Elsevier Online Social Networks and Media (2021). DOI: \url{https://doi.org/10.1016/j.osnem.2021.100155}
\end{textblock*}

\section{Introduction}

Social media applications are a major forum for people to express their opinions and information. By interacting with such applications, users build complex networks that favor the dissemination of information~\cite{Al-Garadi:2018}. Indeed, social media has become an important source of information for a large fraction of the world population~\cite{Shearer:2018, Statista:2020, Newman:2019Reuters}. It has been shown to play an important role in social mobilization and political engagement
~\cite{Resende:2019, Munoz:2017}, notably during major political events~\cite{Pierri:2020}.

Instagram has observed a surge in popularity in recent years~\cite{Whashington:2020}, particularly among the youth. The use of Instagram for consuming news has doubled since 2018, and the platform is set to overtake Twitter as a news source~\cite{Newman-Reuters-b:2020}. Not a surprise, political personalities are increasingly leveraging Instagram to reach the population at scale. Understanding how users interact with each other is paramount to uncover how information is disseminated in the platform and how the online debate impacts our society~\cite{Conover:2012, Gorkovenko:2017, Pierri:2020, Tanase:2020, Resende:2019, Alizadeh:2019, Gorrell:2020}. Prior studies of user behavior on Instagram mainly focused on user engagement based on content type~\cite{Reece:2017, Jaakonm:2017,  Garret:2019, Kao:2019, Kim:2020, Weerasinghe:2020}, general characteristics of comments associated with political messages~\cite{Trevisan:2019, Zarei:2019} and the impact of the posted content on marketing contexts~\cite{Jaakonm:2017, Yang:2019, Kang:2020}. The literature lacks an investigation of the networks that emerge from users' interactions, in particular around political contents, networks that play key roles in information dissemination.

In Instagram jargon, a \emph{profile} is followed by a set of \emph{followers}. A profile with a large number of followers is called an \emph{influencer}. Influencers post content, i.e., \emph{posts}, containing a photo or a video. Followers or any registered user in the case of public profiles can view the profile's posts and comment on them, becoming \emph{commenters}. We here refer to users who comment on the same post as \emph{co-commenters}, and to the interactions that occur among multiple users (often more than two) when they comment on the same post as {\it co-interactions}.  
Co-commenters may form {\it communities} that arise either naturally (i.e., based on common interests on specific topics) or driven by hidden efforts (e.g., ad-campaign or coordinated behavior). By feeding the discussions, communities may favor the spread of specific ideas or opinions while also contributing to increase the visibility of particular influencers. Thus, revealing how such communities emerge and evolve over time is key to understanding information dissemination on the system. 

Studying communities of co-commenters is however challenging. First, users may become co-commenters incidentally because of the high popularity of some posts and/or influencers. Equally, very active commenters naturally become co-commenters of many other users as a side effect of their great frequency of commenting activity.  All these cases are expected to happen, especially given the frequent heavy tail nature of content and user popularity in social media \cite{Ahn:2007}. Conversely, we are interested in finding the true group behavior, driven by users' interests or peer influence~\cite{Burke:2009, Wilson:2009, Kwon:2014}.

Moreover, often happening in large volumes, those incidental co-interactions may lead to the formation of networks of co-interactions with lots of sporadic, uninteresting or weak edges. Investigating the aforementioned communities by looking at the entire networks may be quite misleading, as a lot of sporadic and weak edges may mask the actual group behavior. In contrast, we want to focus on the underlying strong topological structure composed of edges representing {\it salient} co-interactions,\footnote{We use the terms salient co-interactions and salient edges interchangeably.} that is, co-interactions that cannot be rooted in independent users' behavior. We here refer to such structure as the {\it network backbone}. Uncovering the network backbone as well as the communities that compose it and investigating how they evolve over time are thus important steps to understand the dynamics of the online debate, ultimately shedding light on factors driving information dissemination on Instagram.

In this paper, we face the aforementioned challenges to investigate the structural and temporal dynamics of communities formed around salient co-interactions of users on Instagram. Our ultimate goal is to understand the important properties of these communities, with a focus on political content. We model co-commenters' activity as a network where nodes represent commenters and edge weights indicate the number of posts on which both users commented. To filter out uninteresting edges and reveal the underlying network backbone, we employ a reference probabilistic network model in which edges are built based on the assumption that commenters behave independently from each other. Our model takes into account mainly two factors: the popularity of posts, and commenters' engagement towards each influencer. By contrasting the network observed in real data with our reference model, we filter out edges whose weights are within the expected range under the assumption of independent users' behavior, thus uncovering the network backbone representing the group behavior we are interested in. Next, we extract communities using the Louvain algorithm \cite{Blondel_2008}, that we characterize in terms of topological structure, textual properties of comments and discussions carried out, highlighting how community membership and topics of discussion evolve over time. 

We build our study on a large dataset of Instagram comments, including approximately $1.8$ $million$ unique commenters on $36\,824$ posts by $320$ influencers in two countries (Brazil and Italy). The dataset covers two months surrounding major elections that took place in each country. For each country, we selected popular political figures as well as top influencers in other categories (e.g., athletes, celebrities, musicians), aiming to identify differences that characterize the political discussions from other (general) discussions.  We track all influencers' posts, recording all comments (and commenters) associated with those posts. We study each week in isolation to construct consecutive temporal snapshots of co-commenter networks and observe how communities evolve over time.

In sum, in this paper, we tackle three research questions (RQs), which lead to new contributions as follows: 

\begin{itemize}[leftmargin=*]
\setlength{\itemindent}{0em}

\item[]\textbf{RQ1:} What are the characteristics of the network backbones emerging from \emph{salient co-interactions} on Instagram?

By extracting the network backbones, we show that the salient edges of the co-commenters network build better-defined communities than those found in the original (complete) networks, highlighting the need to uncover the network backbone before studying such communities. More specifically, the communities have proportionally more nodes, more edges and are more dense, indicating a more correlated behavior among commenters of the same community.

% In particular, we found higher percentages of salient interactions in politics -- 84\% (87\%) for Brazil (Italy) -- than in other categories -- 65\% (60\%) for Brazil (Italy). These results suggest that influencers in politics can mobilize more commenters, even if they usually post less and are followed by smaller number people than other influencers.

\item[]\textbf{RQ2:} What are the distinguishing properties of the communities that compose such backbones, notably communities formed around political content? 

We characterize the properties of the communities and find that, in politics, multiple communities engage with the same influencer but are typically driven by different topics. Communities in politics tend to be more engaged in online discussions than non-political counterparts. Comments in politics tend to be longer, richer in emojis, hashtags and uppercase words (indicating assertive and emotional content), and tend to carry a more negative tone than other categories.

\item[]\textbf{RQ3:} How do community properties evolve over time?

Aiming to shed light on how information propagates on Instagram through co-commenters' networks, we notice a heating up in political debate and large variations on community membership in weeks preceding elections. Both are reduced after the election day. Yet, top commenters remain consistently active and preserve their community memberships over successive weeks. Moreover, we observe a great diversity in discussed topics over time for communities in politics. Whereas some topics attract attention only momentarily (e.g., racism), others, centered around more fundamental political subjects (e.g., rallies, particular candidates and political ideologies), remain consistently active.

\end{itemize}

This study extends our preliminary work \cite{Ferreira:2020}, where we discussed the emergence of communities around political discussions on Instagram. We extend it by providing a much broader characterization of the communities, covering not only basic topological and temporal properties but also textual properties of the content posted by communities, e.g., topics, sentiment and psycholinguistic properties. As such, we offer a deep analysis of political discussions on Instagram, which helps to understand content propagation during electoral periods.

The remainder of this article is organized as follows:  
Section~\ref{sec:related} summarizes related work, while
Section~\ref{sec:methodology} presents our methodology to extract the  network backbone and analyze co-commenter communities. Section~\ref{sec:dataset} describes our dataset, and Sections~\ref{sec:rq1}-\ref{sec:rq3} present our main results. Finally, Section~\ref{sec:discussion} discusses our findings, offers conclusions and  directions for future work. 

\section{Related work}
\label{sec:related}

We characterize political discussions on Instagram by analyzing user interactions from a network perspective. Three bodies of work are related to our effort: i) studies on online discussions in social media; ii) efforts to model interactions among groups of users (i.e., co-interactions); and iii) methods to extract network backbones.

\subsection{Online discussions in social media applications}

Several social media applications have been largely studied as platforms for political debate. For example, Nguyen ~\cite{Nguyen:2018}  presented a literature review of the role of Twitter on politics, notably as a platform to help politicians win elections and foster political polarization. Indeed, many studies have already argued for the increasing polarization in political orientation~\cite{Gruzd:2014, Vergeer:2015}, whereas others have explored the benefits that politicians can have from using Twitter to reach their supporters~\cite{Chi:2010}. 

Gorkovenko et al.~\cite{Gorkovenko:2017} studied user behavior on Twitter during live political debates, finding that people often use the platform to share their opinions, make provocative  statements, and inform others. Badawy et al.~\cite{Badawy:2018}, in turn,  found evidence of the use of Twitter for political manipulation. In a different direction, Caetano et al.~\cite{Caetano:2018}  identified four groups of  politically engaged users on Twitter during the 2016 US presidential campaign, namely advocates for both main candidates, bots and regular users, characterizing properties of each group.  

Online discussions have also been studied in the context of other platforms. For example, Tanase et al. \cite{Tanase:2020} studied the political debate around the \textit{Brexit} on Facebook, analyzing messages that generated higher engagement from users. In~\cite{fabricio}, the authors developed a system to detect political ads on Facebook and used it to present evidence of misuse during the Brazilian 2018 elections. 
WhatsApp has also been analyzed as an important platform for political debate and 
information dissemination, notably for the spread of fake news during political elections \cite{Resende:2019, Josemar:2019}. 

Considering Instagram, in particular, the literature on user behavior and interactions is reasonably recent and somewhat restricted. Some authors have analyzed user engagement~\cite{Kang:2020, Yang:2019, Kim:2020, trevisan2021debate} and properties of the textual content shared by Instagram users~\cite{Zhan:2018, Arslan:2019}, but with no particular focus on political discussions. Closer to our present effort,  Zarei et al.~\cite{Zarei:2019} analyzed user engagement of twelve Instagram profiles, including profiles of politicians, searching for \emph{impersonators} -- i.e., users who simulate others' behavior to perform specific activities, such as spreading fake news.   Mu\~{n}oz et al.~\cite{Munoz:2017} studied image content posted  by candidates during the US 2016 primary elections, highlighting combined factors that attract user engagement, whereas Trevisan et al.~\cite{Trevisan:2019} performed a quantitative study of the political debate on Instagram, highlighting that politicians' profiles tend to have significantly more interactions than others. 

To the best of our knowledge, we are the first to analyze political discussions on Instagram from a network perspective. As such, we complement those  previous efforts by providing an orthogonal and broader analysis of the dynamics of communities of co-commenters who engage in and  drive political discussions. Our present effort builds on a preliminary work~\cite{Ferreira:2020} where we focused on identifying the structure emerging from the co-commenter network, offering a novel methodology for backbone extraction in this kind of network.  We here greatly extend our prior effort by offering a much more thorough characterization of the content and temporal properties of communities extracted from the backbones.

\subsection{Modeling interactions among groups of users}

A plethora of phenomena related to in-person and online human interactions have been analyzed using concepts of complex networks. Many of these interactions occur among two or more entities simultaneously -- referred here as {\it co-interactions}. Co-interactions constitute an important substructure for some complex systems, exhibiting a slew of properties relevant to the study of the (often global) phenomenon of interest \cite{Benson2:2018, Meng:2018}. %Examples of such co-interactions are co-authors of scientific publications, people co-visiting the same places driven by cultural interests, groups of users commenting or sharing the same topic on a social media application \cite{Benson:2018, Nobre:2020, Giglietto:2020}. 

Some recent studies have empirically analyzed co-interactions through the lens of {\it higher-order networks}, including \textit{motif} (or graphlet) representations, \textit{multilayer} networks, \textit{simplices} and \textit{hypergraphs}~\cite{Meng:2018, Benson:2018, Liu:2019}. In their most basic form, these structures are represented by different motifs (e.g., triangular motifs, star, structural hubs, etc)~\cite{Benson:2016b, Rossi:2018}. Some studies focused on characterizing many of these networks \cite{Zhao:2010,Benson2:2018}, while others analyzed the relationship between motifs and some specific phenomena of interest on a particular network \cite{Adamic:2008,Kuvsen:2020}. Focusing on revealing communities emerging from co-interactions, which is our present goal, some approaches are concentrated at uncovering communities that are based on particular structural patterns, specifically, on motifs \cite{Pizzuti:2017, Yin:2017, Tsourakakis:2017, Huang:2019}.

Conversely, we here aim at uncovering communities of co-commenters who may be driving the online discussions. These communities are not necessarily specific structural patterns, but rather tightly connected subgraphs with respect to the rest of the network. Hence, we focus on the densest and most uniform flavor of co-interaction, where all individuals interact with each other in a motif known as clique~\cite{Battiston:2020}. %We intend to look for communities from the network that emerge from a sequence of such co-interactions driven by actions or interest of multiple individuals.
To do so, we model such co-interactions by projecting them into a weighted and undirected graph, similar to other works present in the literature. For example, Giglietto et al.~\cite{Giglietto:2020} analyzed posts on Facebook focusing on identifying inauthentic behavior by modeling the network of link co-sharing, formed by entities (pages, groups, and verified public profiles) which shared the same content frequently in a short period of time. Cruickshank et al.~\cite{Cruickshank:2020} analyzed interactions among topics on Twitter by modeling a sequence of networks from co-occurrences of hashtags used by Tweeter users. Aiming to reveal coordinated behavior on Twitter, Pacheco et al.~\cite{Pacheco:2020} proposed a set of network models that capture different patterns of co-interactions among users. Examples of patterns include using similar hashtags, sharing the same images or chronological use of the platform. The authors focus on the top 1\% edges as being the most representative. Other works adopt broader definitions of interaction, e.g., users co-retweeting, using the same hashtags or URLs, mentioning the same accounts, or joining the same \textit{conversation} when reply chains with a common root tweet at the same time \cite{Hanteer:2018, Weber:2020}. Yet, they still apply a fixed weight threshold across all edges.

% Giglietto:2020, Pacheco:2020, Nobre:2020, Hanteer:2018, Weber:2020 -> Co-interações com backbone

% Cruickshank:2020, Pacheco, Nobre, - > Community but without backbone.

Thus, as part of our proposal to model co-interactions that occur among Instagram users as they comment on the same post, we must tackle the challenges of using the projected network, notably the presence of a potentially large number of weak and possibly irrelevant edges. Specifically, we adopt an approach that reveals edges in the projected network that, in fact, unveil how the discussion takes place on Instagram. In contrast to prior work \cite{Giglietto:2020, Pacheco:2020, Nobre:2020, Hanteer:2018, Weber:2020}, we remove those co-interactions formed by chance, due to the frequent heavy tail nature of the content and user popularity in social media \cite{Ahn:2007}. To address this challenge, we propose a generative model to filter such noisy edges out of the network, thus retaining only {\it salient} edges in the network backbone. In the next section, we review prior works focused on this task, distinguishing our method from existing alternatives.

% we investigate strategies that remove such noise and reveal edges in our co-commenters network that, in fact, contribute to the study of information dissemination from the co-interaction of users on Instagram.  Moreover, we investigate strategies to extract the backbone in such context and discuss their limitations in the next section shedding light into a new approach for address such challenge.

% use a generative model to filter such noisy edges out of the network, thus retaining only {\it salient} edges in the network backbone.  In the next section, we briefly review prior work on network backbone extraction, distinguishing the method we employ from existing alternatives. 

% In contrast to prior work, we  model the co-interactions that occur among Instagram users as they comment on the
% same post by projecting them into a set of pairwise interactions. However, as already argued \cite{Liebig:2016, Benson:2018, Coscia:2019, Cao:2019,  Kumar:2020}, such projected networks may include a large number of random, sporadic and weak edges that are not really part of the fundamental underlying network backbone. Rather they occur mostly by chance or as a result of independent user behavior.  Thus, we use a generative model to filter such noisy edges out of the network, thus retaining only {\it salient} edges in the network backbone.  In the next section, we briefly review prior work on network backbone extraction, distinguishing the method we employ from existing alternatives.

\subsection{Network backbone extraction}

Network backbone extraction consists of removing edges of a network to retain a clearer image of fundamental structures governing the network~\cite{Coscia:2017}. As  several network models and algorithms assume  that edges faithfully represent the interactions under study, backbone extraction is a necessary step to avoid that spurious, weak and sporadic edges blur and impair the investigation~\cite{Slater:2009, Newman:2018}.

Some of the simplest methods of backbone extraction explore topological metrics and global properties of the graph. They typically propose to select the most important edges for the study \cite{Leao2018, Hanteer:2018, Pacheco:2020, Giglietto:2020, Nobre:2020}. Examples include the use of K-core searches for dense subgraphs~\cite{Sariyuce:2016, Savic:2019} and the removal of edges based on a global threshold $\tau$, either applied directly to edge weights~\cite{Ferreira:2018, Namaki:2011, Yan:2018}  or to more sophisticated metrics such as the neighborhood overlap of a pair of nodes~\cite{Brandao:2017, Ferreira:2019}. These methods are particularly adequate when the concept of salient edge is well-defined by the problem context, e.g., large edge weight \cite{Ferreira:2018, Yan:2018}.  However, threshold-based approaches may lead to misleading interpretations and introduce bias in the analyzes, since setting the $\tau$ threshold appropriately depends on the context and therefore can be quite complex \cite{Tsur:2017}. 

Other methods are based on local probability distributions, i.e., distributions specific to each edge or subset of edges. For instance, Serrano et al.~\cite{Serrano:2009} propose the \textit{disparity filter} method, based on the assumption that an edge is salient if two nodes are connected with a disproportionately high weight compared to the weights connecting them to their other neighbors. In other words, salient edges are those that have weights that deviate significantly from the null hypothesis that the weights of all edges incident to a given node are uniformly distributed. Edge weights are compared to the reference model %Thus, edges are expressed as parts of the total edge weights of a node. These parts are then compared to a reference model,
and $p$-values are used to determine how much an edge differs from it.
\textit{Noise Corrected} \cite{Coscia:2017} is another local backbone extraction method, based on the assumption that the salience of an edge results from collaboration between the nodes. Unlike the disparity filter, this method is able to preserve peripheral-peripheral connections by estimating the expectation and variance of edge weights using a binomial distribution. It considers the propensity of the origin node and the destination node to emit and receive edges. To be regarded as salient, an edge must exceed its expected weight considering the connection strength of its nodes. 

%Another probabilistic method was proposed by Grady et al. \cite{Grady:2012}. It extracts the network backbone through a sampling process which entails calculating the tree of shortest paths connecting the given node to all other nodes on the network. Then, the distribution of the number of times that an edge appears in these shortest paths is used to determine whether a given edge is salient. Note that if an edge was created at random but still plays an important structural role in the network, it's likely to be considered salient by this method. % This approach has been adopted to studying on phenomena related to interactions between species (for biological purposes) \cite{Ma:2019} and analyzing the properties of co-authorship networks from the perspective of influential nodes \cite{Petri:2013}. 
% Other approaches focus on specific domains (e.g, recommendation systems, urban mobility, wireless sensor networks) prune nodes using contextual information of the domain-specific reference model~\cite{Zeng:2016, Manca:2017, Kumar:2020}.

Finally, other methods to extract the network backbone make use of a reference model  describing how the network should be built under certain assumptions -- e.g., interactions occur by chance, based on independent behavior \cite{Jacobs:2015} or uniform weight distribution over edges incident to a node~\cite{Slater:2009}. The general idea is to keep only edges that deviate enough (in a probabilistic sense) from these assumptions. For example, under the assumption of independent behavior,  the authors of~\cite{Silva:2014} propose to filter out edges whose weights are below the $95^{th}$ percentile of the edge weight distribution defined by the reference model. 

We here employ a generative model to extract the backbones from co-commenters networks.
This method was originally proposed in a preliminary version of this work \cite{Ferreira:2019}. In contrast to the aforementioned methods, our model considers the fundamental elements and structure of the target domain (influencers, commenters and posts). Specifically, it takes into account \textit{post popularity} and \textit{user activity level} as factors to model independent user behavior on a per-edge basis, keeping only edges that deviate significantly from it. As such, our method relies on a fine-grained local reference model to keep only salient edges, thus revealing the network backbone.

\section{Methodology}
\label{sec:methodology}

In this section we formally define the network of co-commenters on Instagram and describe the probabilistic network model used as reference to uncover salient interactions. We then describe how we extract communities from the network backbone and present the techniques employed to characterize these communities.% temporal dynamics and textual properties of posted content. 
\subsection{Network of co-commenters}
\label{sec:netw}

We model the dynamics of interactions among users who comment on the same  Instagram post as a sequence of snapshots  of fixed time window $w$, where each snapshot aggregates posts of a selected set of influencers and their associated comments. We here consider $w$ equal to one week as a reasonable period to cover discussions around posts.

Given a time window $w$, we take the corresponding set of posts $P_w$, whose creation times fall within $w$,  to create a weighted and undirected graph $G_w=(V_w,E_w)$. Vertices in set $V_w$ correspond to users who commented in at least two posts in $P_w$. We choose to disregard commenters whose activities were concentrated on a single post, and thus reflect sporadic behavior.\footnote{Note that, by doing so, commenters who commented multiple times on a {\it single} post, but did not comment on other posts, are removed.} On Instagram, commenters can also reply directly to another user's comment. In our problem, we are interested in finding users with similar behavior or interested in a similar topic or influencer. As such, we build our network uniquely based on the appearance of commenters in a post's comments and neglect whether they are answering to previous comments. This is also supported by our observation that, in many cases, commenters engage in discussion on specific topics without using the \emph{reply to comment} feature. Let $P_w(c) \subseteq P_w$ be the set of posts on which user $c$ commented. An edge $e_{cd} = (c,d)$ is added to set $E_w$ if $P_w(c) \cap P_w(d) \neq \emptyset$, i.e.,  commenters $c$ and $d$ commented at least once on the same post. Thus, edges link co-commenters. The weight $\gamma({cd})$ of edge $e_{cd}$ is defined as the number of posts on which $c$ and $d$ commented together, i.e., $\gamma({cd}) =  |P_w(c) \cap P_w(d)| \in \{1,2,\ldots,|P_w|\}$. Aiming at characterizing similarities and differences in discussions across different themes (political or not) and countries (Brazil and Italy), we build separate networks $\mathcal{G}=\{G_1, G_2 ... G_n\}$, where $n$ is the total number of time windows $w$,  for each scenario.

By definition, each post $p$ in $P_w$ generates a clique (i.e., a complete subgraph) in the graph.
%, i.e., a subgraph where all pairs of vertices are adjacent. In other words, each subgraph induced by a single post $p$ is complete.
Thus, the final network $G_w$ is the superposition of all the cliques generated by the posts in $P_w$. As such, $G_w$ results in a complex network with a large number of vertices and edges, many of which may be the result of independent behavior of different commenters, i.e., not a reflection of actual discussions. For example, a very popular post leads to a large clique in the graph. Yet, many co-interactions captured by this clique are likely a side effect of the popularity of the post, or of the influencer who created it. Similarly, a user who is very active will likely co-occur as a commenter with many others. That is, such co-interactions are to some extent {\it expected} given users' activity and post popularity. Thus, to analyze interactions among co-commenters, we filter out such expected edges and focus on those whose frequencies of occurrence are large enough to allow us reject, with some confidence,  the assumption of independent behavior. That is, we focus on salient edges that most probably reflect real online discussions, forming the underlying fundamental network backbone. 
%compose the fundamental underlying structure, i.e., the \emph{network backbone}.

\begin{figure*}[!t]
    \centering
    \subfloat[Raw Data]{\includegraphics[width=1\columnwidth]{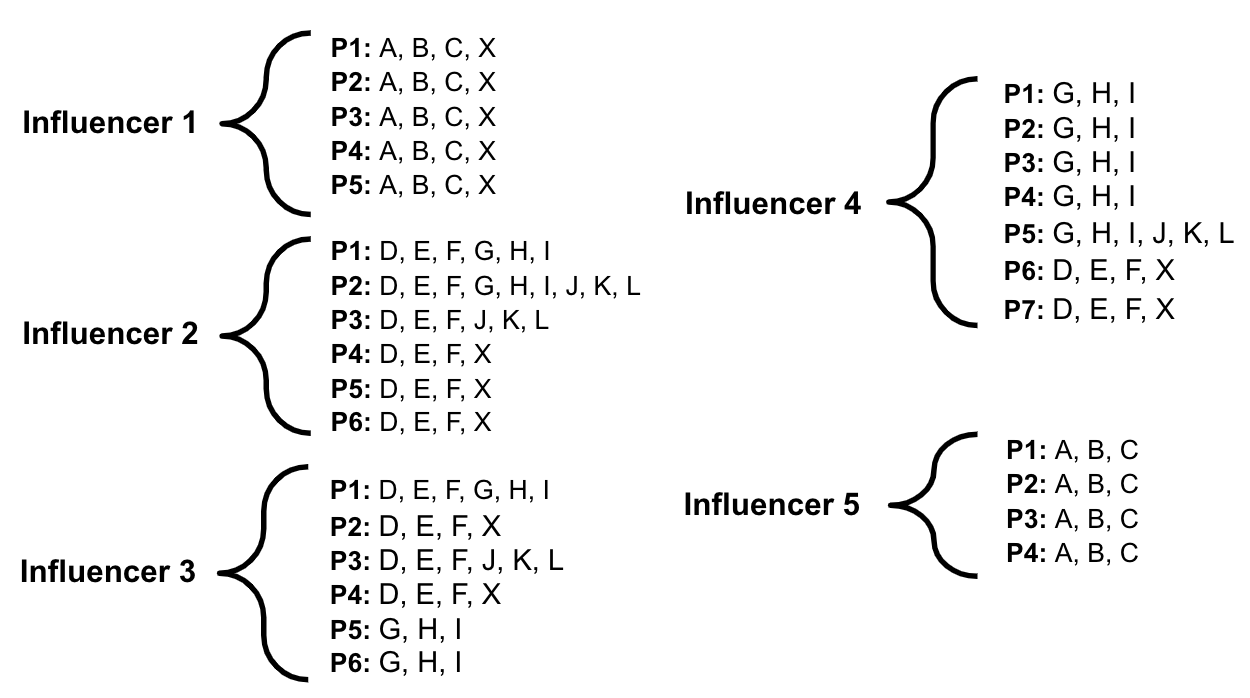}\label{Fig:toy_fig_1}}\hspace{0.32cm}
    \subfloat[Original Network $G_w$]{\includegraphics[width=0.45\columnwidth]{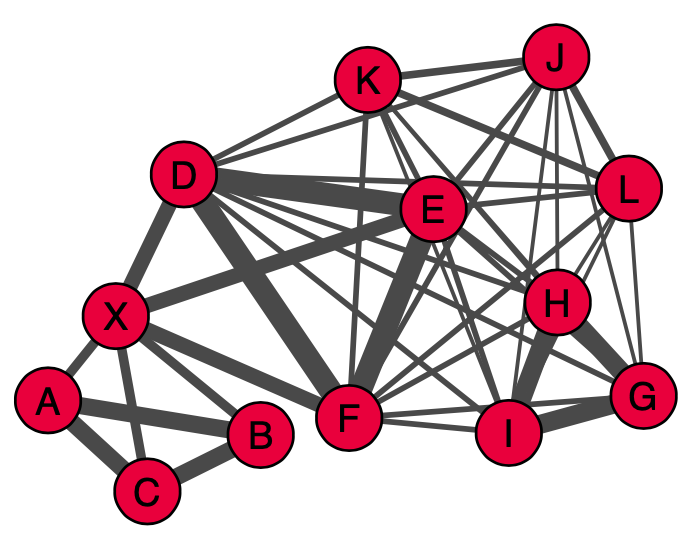}\label{Fig:toy_fig_2}\vspace{1.4cm}}\hspace{0.4cm}
    \subfloat[Network Backbone $B_w$]{\includegraphics[width=0.42\columnwidth]{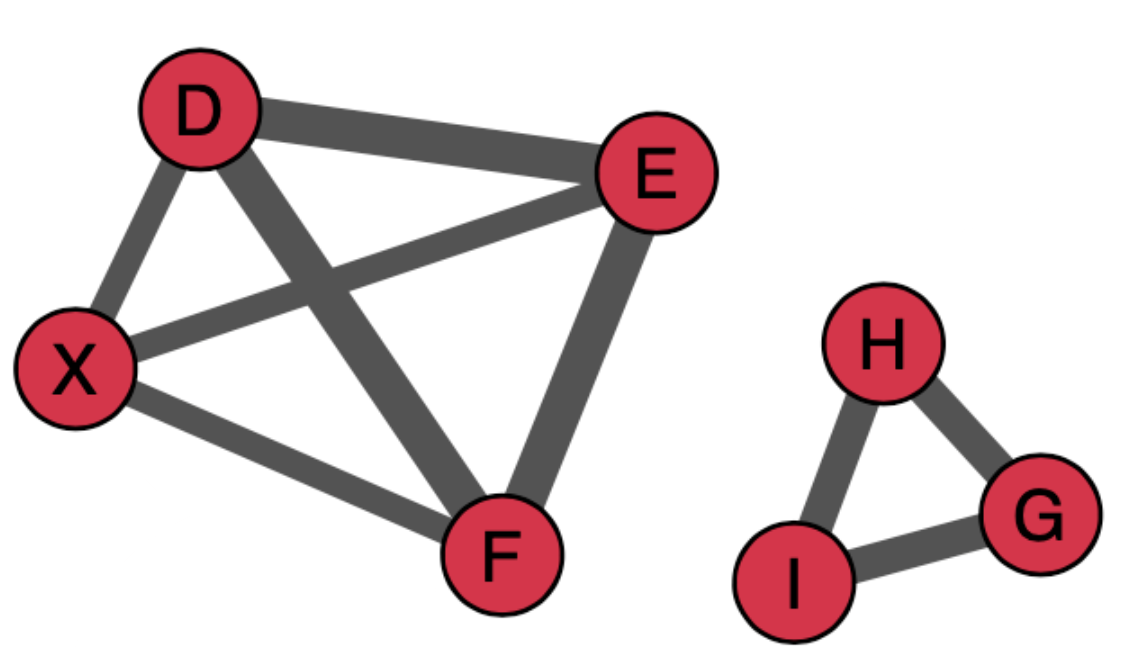}\label{Fig:toy_fig_3}\vspace{1.4cm}}
    \caption{Illustration of the backbone extraction process in a simplistic graph. The isolated vertices are removed from the final $B_w$ used in our analysis.}
    \label{Fig:toy_fig}
\end{figure*}

\subsection{Network backbone extraction}
\label{sec:back}

A fundamental question that arises when studying complex networks is how to quantify the statistical significance of an observed network property~\cite{Coscia:2017, Newman:2018}. To that end, reference models are often used to determine whether networks display certain features to a greater extent than expected under a null hypothesis (e.g., independent behavior)~\cite{Burke:2009, Wilson:2009, Kwon:2014}. A reference (or {\it null}) model matches some of the features of a graph and satisfies a collection of constraints, but is otherwise taken to be an unbiased random structure. It is used as a baseline to verify whether the object in question displays some non-trivial features (i.e., features that would not be observed as a consequence of the constraints assumed). An appropriate reference model behaves according to a reasonable null hypothesis for the behavior of the system under investigation. One strategy to build a reference model is by employing  generative growing networks \cite{Easley:2010, Newman:2018}.

We here employ a reference generative model $\widehat{G}_w$  for each network $G_w$ that is based on the hypothesis that commenters during $w$ behave independently from each other~\cite{Ferreira:2020}.  That is, edge weights in $\widehat{G}_w$  are defined under a generative process in which commenters act independently from each other, although their interactions with influencers' posts (i.e., which post each user comments on)  are not identically distributed.
We can then observe which edges of the real network $G_w$ do not behave in accordance with the reference model $\widehat{G}_w$ -- i.e., reflect interactions that significantly deviate from an independent behavior. Such edges will compose the network {\it backbone}. Intuitively, we want to highlight co-interactions that occurred more often than what would be expected if commenters behaved independently.

Our null model takes as input the  popularity of each post (number of unique commenters) and the engagement of commenters towards each influencer (number of posts by the influencer each commenter writes on). Using these statistics, comments are randomly assigned to commenters while preserving: i) the set of influencers on which each commenter writes a comment; ii) the popularity of each  post, and iii) the engagement of each commenter towards each influencer. The model assigns commenters to each post using independent and identically distributed (i.i.d.) draws from a distribution where the probability is proportional to the commenter's engagement towards the target influencer.   By doing so, we prevent the backbone from being dominated by very active commenters or by those engaged in highly popular posts.

More specifically, let $I_w$ be the set of all influencers who wrote posts in $P_w$. Let $\mathcal{C}_p \subseteq V_w$ be the set of unique commenters in post $p \in P_w$ and $\{\mathcal{{P}}_w^i\}_{i \in I_w}$ be a partitioning of $P_w$ based on the influencer
$i \in I_w$ who created the post. We define the engagement of commenter $c \in V_w$ towards influencer $i$ (measured by the total number of posts in $\mathcal{P}_w^i$ commented by $c$) as
\begin{equation}
    x_i(c) = \sum_{p \in \mathcal{P}_w^i} \mathds{1}\{c \in \mathcal{C}_p\},
\end{equation}
where $\mathds{1}\{.\}$ is the identity function. We then  define $c$'s \textit{relative engagement} towards $i$ w.r.t.\ other commenters as:
\begin{equation}
f_i(c) = \frac{x_i(c)}{\sum_{d \in V_w} x_i(d)} = \frac{x_i(c)}{\sum_{p \in \mathcal{P}_w^i} |\mathcal{C}_p|}.
\end{equation}

In this way, we can describe in details the three steps of the generative process to build our reference model~$\widehat{G}_w$:

\begin{enumerate}
    \item For each post $p \in P_w$, we consider a random assignment of each of the $|\mathcal{C}_p|$ (unique) commenters  to a commenter $c \in V_w$ with probability $f_i(c)$, where $i$ is the author of $p$. Specifically, under the assumption of independent behavior, we consider each such assignment as a Bernoulli random variable with parameter $f_i(c)$.  The probability that commenter $c$ is not assigned to $p$ is thus a Binomial random variable, with 0 successes in $|C_p|$ experiments. Conversely, under the assumption of independent behavior,  the probability that $c$ has commented (at least once)  on a post $p \in \mathcal{P}_i$ is $r_p(c)=1-(1-f_i(c))^{|\mathcal{C}_p|}$.
    \item For each pair of commenters $c$ and $d$, we denote by $r_p(c,d)$ the probability that both get assigned to post $p$ and by $r_p(d|c)$ the probability that $d$ gets assigned to $p$ given that $c$ is assigned to $p$.  The conditional probability $r_p(d|c)$ is necessary because, strictly speaking, although we are drawing commenters independently, when $c$ is drawn, it decreases the number of chances $d$ has for being drawn (since $|\mathcal{C}_p|$ is fixed). Hence, 
    $r_p(c,d) =   r_p(c) \times r_p(d|c).$
    We approximate $r_p(c,d)\approx r_p(c)\times r_p(d)$, for each $p \in P$. Intuitively, this approximation works well when $|\mathcal{C}_p|$ is large (as in the case of most influencers' posts), because drawing $c$ decreases by only one the number of draws that can be used to draw $d$. Then, for each post $p\in P_w$, our model defines a distribution  over the set of vertices corresponding to  $\mathcal{C}_p$, where the value of the random variable $\widehat{\Gamma}_p{(c,d)} \in \{0,1\}$ indicates the existence of an edge between commenters $c$ and $d$, and is given by a Bernoulli trial with parameter $r_p(c,d)$, i.e.\ $\widehat{\Gamma}_p{(c,d)} \sim \textrm{Bernoulli} (r_p(c,d))$. 
     \item The reference model $\widehat{G}_w=(\widehat{V}_w, \widehat{E}_w)$ is composed by the superposition of all the edges created for all $p \in P_w$. Hence, an edge $\widehat{e}_{cd} \in \widehat{E}_w$ will have a weight distribution described by a random variable $\widehat{\Gamma}{(c,d)} =  \sum_{p \in P_w} \widehat{\Gamma}_p{(c,d)}$. Therefore, it will be a sum of Bernoulli random variables with distinct probabilities \cite{Wang:1993}, which follows a Poisson Binomial distribution with parameters  $ r_1(c,d), r_2(c,d),\ldots,$ $ r_{|P_w|}(c,d) $.
\end{enumerate}

We can then compare the reference model $\widehat{G}_w$ with the observed network ${G}_w$ to extract the backbone $B_w$ of the latter. We do so by keeping in $B_w$ only edges of ${G}_w$ whose weights have values exceeding the ones expected in $\widehat{G}_w$ by a large margin. Specifically, for each edge $\widehat{e}_{cd}$ we compute the $(1-\alpha)^{th}$ percentile, denoted by $\widehat{\gamma}_{1-\alpha}(c,d)$, of the distribution of edge weight $\widehat{\Gamma}(c,d)$, and compare it with the observed edge weight ${\gamma}{(c,d)}$. We keep edge $e_{cd}$  if ${\gamma}{(c,d)} > \widehat{\gamma}_{1-\alpha}{(c,d)}$. Intuitively, we keep only edges between co-commenters who interacted much more often than expected under the assumption of independent behavior. That is, edges for which the  chance of such frequency of interactions  being observed under the independence assumption is below $\alpha$. We here set $\alpha$ = 5\%, as done in prior studies \cite{Serrano:2009, Kobayashi:2019}. Note that, after filtering out edges, isolated vertices are also removed. At the end, we extract from the network $G_w$ its backbone $B_w= (V^b_w, E^b_w)$ where $V^b_w \subseteq V_w$ and $E^b_w \subseteq E_w$.

Note that the $(1-\alpha)^\textrm{th}$ percentile is computed separately for each edge $e_{cd} \in E_w$ from  random variable $\widehat{\Gamma}(c,d)$. For such a Poisson binomial distribution, there is a closed form for computing a given percentile~\cite{Hong:2013}, which, however, is expensive to compute. Instead, we here use the Refined Normal Approximation (RNA)~\cite{Hong:2013}, a method that proved very good performance with low computational complexity.

\subsection{Backbone extraction exemplified}

We illustrate how the backbone is extracted from a given input network $G_w$ by means of the toy example shown in Figure~\ref{Fig:toy_fig}. Figure~\ref{Fig:toy_fig_1} shows a total  of five influencers, each with a different number of posts ($P1$, $P2$, etc), and each post with associated commenters ($A$, $B$, etc). Posts have different popularity, and commenters have different activity levels and engagement towards each influencer. The projected graph $G_w$ is depicted in Figure \ref{Fig:toy_fig_2}, whereas the extracted backbone $B_w$ is shown in Figure \ref{Fig:toy_fig_3}. In both networks, line thickness is proportional to the edge weight. 

The question that arises is: {\it  why did we extract only the edges shown in Figure \ref{Fig:toy_fig_3} to compose the network backbone?} Recall that our model selects as salient edges those that have weights large enough so that we can reject the assumption of independent user behavior.   Thus, for each edge in $G_w$, we ask ourselves: is there enough evidence to reject the assumption of independent behavior? If so, the edge is kept; otherwise, it is removed. 

Let’s illustrate our decisions regarding four groups of edges, focusing first on edges incident to commenters $A$, $B$, $C$. Note that all three commenters commented on posts only by influencers  $1$ and $5$ and they commented on {\it all} posts by both influencers.  These commenters are thus quite active, and the popularity of these posts is actually high, considering the population of users who commented on them. As such,  it is possible that $A$, $B$ and $C$ are driven by their individual interests on these two influencers, and, as such, most probably would comment in most (if not all) posts by them. Thus, based on the observed data, we cannot reject the assumption of independent user behavior when considering co-interactions among $A$, $B$ and $C$ and the corresponding edges are not kept as part of the network backbone in Figure ~\ref{Fig:toy_fig_3}.  For example, the edge $e_{AB}$ has weight ${\gamma}{(A,B)} = 9$ which is below or equal to the $95^{th}$ percentile of the corresponding edge weight distribution $\widehat{\gamma}_{0.95}{(A,B)} = 9$.  The same reasoning applies to commenter $X$, who only commented on posts by influencer $1$. Thus, the co-interactions  of $X$ with $A$, $B$ and $C$ are {\it not} considered salient and the corresponding edges are not kept.   

Let’s consider now the edges incident to commenters $J$, $K$ and $L$.  These users  co-comment with low frequency  in posts by influencers $2$ ($P2$ and $P3$), $3$ ($P3$) and $4$ ($P5$). These posts are the most popular posts by  such influencers, receiving comments from several other users as well. It is therefore expected that commenters active on these posts will have several co-commenters, as we can observe in  Figure ~\ref{Fig:toy_fig_2}.  However, when it comes to $J$, $K$ and $L$, the weights of these edges are small, as the co-interactions are somewhat sporadic. Moreover, note that the posts on which these users commented are among the most popular ones by the corresponding influencers, attracting most of their commenters. For example, $P2$ by influencer $2$ received comments by 9 out of all 10 users who commented on her posts.  Co-interactions built around such highly popular post are {\it not} considered salient as one cannot tell whether commenters are truly interacting with each other or simply reacting independently to a quite attractive content. From an operational perspective, recall that, when building the reference model $\widehat{G}_w$ we do need to assign commenters to comments associated with each post. In the case of such very popular posts, most if not all potential commenters are assigned, thus raising the chance of the edge being added to $\widehat{G}_w$, and thus of the edge being considered expected under the assumption of independent behavior.

We now turn our attention to the edges incident to two groups of commenters:  i) $D$, $E$, $F$ and $X$; and  ii) $G$, $H$, $I$.  In both cases, the commenters co-interact on posts by influencers $2$, $3$ and $4$, and the co-interactions occur very often on different posts by these influencers. However, unlike  the case of $A$, $B$ and $C$, discussed above, there are other users who also commented on the same posts. Compared to these other commenters,  $D$, $E$, $F$, and $X$ (as well as $G$, $H$, $I$) clearly stand out as frequent co-commenters. That is, taking the overall behavior of the commenters of these posts,  we find that the co-interactions among $D$, $E$, $F$, and $X$ (as well as $G$, $H$, $I$) are more frequent than expected if these users were being driven by independent behavior.  For example,  the weight of edge $e_{DE}$ is  ${\gamma}{(D,E)} = 12$ which is larger than the $95^{th}$ percentile of the corresponding edge weight distribution $\widehat{\gamma}_{0.95}{(D,E)} = 10$.  We consider this evidence strong enough to reject the assumption of independent behavior. The same holds for the other aforementioned commenters.  As consequence, the corresponding edges are maintained in the backbone (see Figure ~\ref{Fig:toy_fig_3}).

Finally, we note that all isolated nodes are removed from the final network backbone (see, for example, nodes $A$, $B$, $C$, $K$, $J$, and $L$, no longer present in Figure~\ref{Fig:toy_fig_3}).

% Com 95 percentil D, E, F limiar é 10, eles aparecem 12
% Com 95 percentil G, H, I limiar é 9, eles aparecem 10
% Com 95 percentil J, K, L limiar é 4, eles aparecem 4

% D, E, F com G, H e I  - qualquer arestas cruzando estes dois grupos -  o limiar é 6 e eles aparecem 3
% J, K, L com G, H e I  - qualquer arestas cruzando estes dois grupos -  o limiar é 5 e eles aparecem 2

\subsection{Community detection}
\label{sec:communities}

Once extracted the backbone $B_w$, our next step consists of identifying communities in $B_w$. Qualitatively, a community is defined as a subset of vertices such that their connections are denser than connections to the rest of the network. To extract communities from $B_w$, we adopt the widely used Louvain algorithm~\cite{Blondel_2008,Newman:2004}. 
The goal of the Louvain algorithm is to maximize the {\it modularity} of the communities. Given the backbone $B_w = (V^b_w, E^b_w)$,  the modularity is defined as:

$$ Q={\frac {1}{2M}}\sum \limits _{c,d \in V^b_w}{\bigg [}\gamma({cd})-{\frac {k({c})k({d})}{2M}}{\bigg ]}\delta (\ell({c}),\ell({d}))$$

\noindent where $\gamma(c,d)$ is edge weight between vertices $c$ and $d$ ($e_{cd} \in E^b_w$);  $k(c)$ and $k(d)$ are the sums of the weights of the edges attached to  $c$ and $d$, respectively, in $B_w$; $M$ is the sum of all  edge weights in $B_w$ ;  $\ell({c})$ and $\ell({d})$ are the communities assigned to $c$ and $d$; and     $\delta({\ell({c}),\ell({d})})=1$ if $\ell({c})=\ell({d})$,  $0$ otherwise.

Intuitively, the modularity captures how much densely connected the vertices within a community are, compared to how connected they would be in a random network with the same degree sequence. Modularity is defined in the range of -0.5 to +1, and modularity scores of 0.5 or higher are considered strong evidence of well-shaped communities. The Louvain method is a heuristic that operates by finding first small communities optimizing modularity locally on all vertices. Then, each small community is merged into one meta-vertex and the first step is repeated.
%Modularity measure the quality of division of a network into  communities. Its values range from -1 to 1 such that a high modularity have dense connections between the nodes within communities and sparse connections between nodes in different communities.
The final number of communities is the result of an optimization procedure. 
We refer the reader to~\cite{Blondel_2008} for a detailed description of the Louvain algorithm.

%In other words, given the graph $B_w = (V^{b}_w, E^{b}_w)$,  the Louvain algorithm extracts the set of communities that provides the highest modularity value. 

\subsection {Community characterization}
Once communities are extracted, we characterize them in terms of the textual properties of the content shared by their members as well as their temporal dynamics.

\subsubsection{Content properties}  

We analyze the discussions carried out by each community by focusing on the textual properties of the comments shared by its members. In particular, we employ three complementary textual analysis approaches. 

First, we perform sentiment analysis using SentiStrength,\footnote{\url{http://sentistrength.wlv.ac.uk/index.html}} a lexical dictionary labeled by humans with multi-language support, including Portuguese and Italian. Given a sentence, SentiStrength classifies its sentiment with a score ranging from -4 (extremely negative) to +4 (extremely positive)~\cite{Thelwall:2010}. SentiStrength has been widely applied to analyze the sentiment of social media content, notably short texts (e.g., tweets), for which identifying sentiment is usually harder~\cite{Ribeiro:2016, Thelwall:2017}. We noticed a high frequency of emojis in the comments. To improve SentiStrenght's ability to recognize them, we incorporate its emoji dictionary with the labeled emoji dataset provided in \cite{Novak:2015}. We also observed the presence of informal writing, slang abbreviations, and wrong work variation could impact our conclusions. Then, we pre-process the comments keeping only comment words with regular words in the country language. To this end, we used Brazilian and Italian dictionaries in Hunspell\footnote{\url{https://github.com/hunspell/hunspell}} format and matched the words found in the comment against them.

Second, we use \emph{Term Frequency - Inverse Document Frequency} (TF-IDF)~\cite{Jones:1972} to reveal terms that characterize each community. TF-IDF is traditionally used to describe \emph{documents} in a collection with their most representative terms. Given a particular term and a document, the TF-IDF is computed as the product of the frequency of the term in the given document ($TF$) and the inverse of the frequency at which the term appears in distinct documents ($IDF$). Whereas $TF$ estimates how well the given term describes the document, $IDF$ captures the term's capacity to discriminate the document from others. 
To apply TF-IDF in our context, we  represent each community as a {\it document} consisting of all comments of the community members.
We pre-process the comments to  %stop words, perform stemming, remove the top-1\% most popular terms, and remove rare words (i.e., less than 10 occurrences). 
remove emojis, stopwords, hashtags, punctuation and mentions to other users,   perform stemming, as well as  remove the overall top-1\% most popular terms and rare terms (less than 10 occurrences).\footnote{The former are words whose frequency is extremely high and would not help to characterize the communities, while  the latter are mostly typing errors or grammar mistakes.}

 Each community is then represented by a vector $d$ with dimension equal to the number of unique terms in the collection. The element $d[i]$ is the TF-IDF of term $i$.
 We here use a modified version of $IDF$, called probabilistic inverse document frequency~\cite{Baeza:1999}, which is more appropriate when the number of documents is small (as is our case). It is  defined as $IDF(i)=\log \frac{N-n_i}{n_i}$, where $N$ is the total number of communities and $n_i$ is the number of communities using the term $i$. 
We manually evaluate the terms with large TF-IDF of each community searching for particular subjects of discussion. 

Last, we delve deeper into community contents using LIWC~\cite{Tausczik:2010}, a lexicon system that categorizes text into psycholinguistic properties. LIWC organizes words of the target language as a hierarchy of categories and subcategories that form the set of LIWC attributes. Examples of attributes include linguistic properties (e.g., articles, nouns and verbs), affect words (e.g., anxiety, anger and sadness) and cognitive attributes (e.g., insight, certainty and discrepancies). The hierarchy is customized for each language, with 64 and 83 attributes for Portuguese and Italian, respectively. We apply LIWC to each comment of each community to quantify the fraction of words that falls into each attribute. We search for statistical differences across communities based on the average frequencies of their respective attributes. We first use Kruskal's non-parametric test  to select only attributes for which there is a significant difference across communities~\cite{Kruskal:1952}. Then, we rank attributes with significant differences to select the most discriminative ones using the Gini Coefficient \cite{Yitzhaki:1979}. 

%When presenting results, we translate all attributes to English, even if they are encoded in the original language in LIWC. We apply LIWC to each comment of each community to quantify the fraction of words in the comment that falls into each attribute. By analyzing these fractions, we can identify attributes that characterize comments of a community. However, not all attributes appear in a comment. In addition, given an attribute, we are interested in knowing whether its frequency is significantly different for comments made by the various communities. 

\subsubsection{Community temporal dynamics}
\label{sec:metho_tempo}

Finally, we analyze how communities evolve over time, in terms of their  memberships as well as main topics of discussion. 

To analyze the dynamics of community membership we use two complementary metrics: persistence and normalized mutual information. 
 \textit{Persistence} captures the extent to which commenters remain in the backbone across consecutive time windows. The persistence at  window $w$+$1$ is given by the fraction of commenters in $B_w$ who are present in $B_{w+1}$. If persistence is equal to 1, all commenters in $w$ are still present in $w$+$1$ (plus eventually others). Yet, the membership of individual communities may have changed as members may switch communities. 

We also use the \emph{Normalized Mutual Information} (NMI)  metric \cite{Shannon:2001} to measure changes in community membership from window $w$ to $w+1$.  Given two sets of partitions $X$ and $Y$ defining community assignments for vertices,  the mutual information of $X$ and $Y$ represents the informational overlap between $X$ and $Y$. Let $P(x)$ be the probability that a vertex picked at random is assigned to community $x$ in $X$, and $P(x,y)$ the probability that a vertex picked at random is assigned to both $x$ in $X$ and $y$ in $Y$. Let $H(X)$ be the Shannon entropy for $X$ defined as $H(X)=-\sum_x P(x) \log P(x)$. The NMI of $X$ and $Y$ is defined as:
\begin{equation}
NMI(X,Y)=\frac{\sum_x \sum_y P(x,y)\log \frac{P(x,y)}{P(x)P(y)}}{\sqrt{H(X)H(Y)}}
\end{equation}

NMI ranges from 0 to 1 where 0 implies that  all commenters changed their communities and  1 implies that all commenters remained in the same community.  We compute NMI at each time window $w$+$1$  by taking the sets of communities identified in windows $w$ and $w$+$1$, considering only members who persisted from the backbone $B_w$ to the backbone $B_{w+1}$.

%and measures the agreement of the two assignments, ignoring permutations. It is widely used for comparing clusterings~\cite{vinh2010information} and community assignments~\cite{danon2005comparing}. NMI ranges from $0$ to $1$ where $0$ indicates that all commenters have changed their communities and $1$ indicates that all commenters remained in the same community. We compute NMI by taking the sets of communities identified in windows $w$ and $w+1$, considering only members who persisted from the backbone $B_w$ to the backbone $B_{w+1}$. We refer the reader to~\cite{danon2005comparing} for the complete formulation of NMI and its applications on graph analysis.

%\ju{In the intro we say that we also analyze other aspects of dynamics. What is correct??? Need consistency here!} 

%\ch{ Aqui 

We also analyze how the topics of discussion of each community evolve over time. To that end, we focus on the most representative  terms used by each community, as captured the the TF-IDF metric,  to investigate to what extent communities use the same lexicon over consecutive time windows. We start by first generating, for each time window, the vector representation of each identified community (as described in the previous section). Given the large size of the vocabulary, we consider only the top-100 words with the highest TF-IDF scores in each document, zero-ing other entries in the TF-IDF vectors. Next,  we need to match the communities found in week $w$+$1$ to the communities found in week $w$ so as to be able to  follow users commenting on the same topics across windows. Rather than doing so by using the structural information, we match them based on the topics or, more precisely, on the set of terms they used in each window.

 Specifically, we use the cosine similarity \cite{Baeza:1999} of the TF-IDF vectors\footnote{The similarity between communities $c_j$ and $c_k$ is defined as  $sim(c_j,c_k) = d_j \times d_k$, where $d_j$ and $d_k$ are the TF-IDF vector representations of communities $c_j$ and $c_k$, respectively. Note that  $sim(c_j,c_k)$ ranges from $0$ (maximum dissimilarity) to $1$ (maximum similarity).} to compute the pairwise similarity between all pairs of communities in windows $w$ and $w$+$1$, matching each community  $c^{w}_j$ in window $w$ with  the most similar one in window $w$+$1$, provided that this similarity exceeds a given criterion of significance.  
 The criterion we adopt consists of comparing the similarity between two communities  $c^{w}_j$ and $c^{w+1}_k$  and the similarity between $c^{w}_j$ and an ``average" community in window $w$+$1$  
Let $\mathbf{d}_j^{w}$ be the TF-IDF vector representation of community $j$ in window $w$, we use {\it all} comments associated with window $w$+$1$ to compute its TF-IDF vector $\mathbf{d}_*^{w+1}$  using the term frequencies in the complete {\it document} (i.e., all comments) but the IDF values previously computed considering individual communities in $w$+$1$.  In practice, the cosine similarity between the TF-IDF vectors $\mathbf{d}_j^{w}$ and $\mathbf{d}_*^{w+1}$ gives us a significance threshold for matching the communities, i.e., when $\textrm{sim}(\mathbf{d}^{w}_j, \mathbf{d}^{w+1}_k) > \textrm{sim}(\mathbf{d}^{w}_j, \mathbf{d}^{w+1}_*)$, the similarity between $c^{w}_j$ and $c_k^{w+1}$ is larger than the similarity between $c^w_j$ and an ``average community'' in window $w+1$. In case no community $c^{w+1}_k$  satisfies that condition, we deem that no match was found for $c^{w}_j$. Instead, if we find a match, it means that we have a significant mapping between two communities in different windows.

\section{Dataset}
\label{sec:dataset}

We now describe the dataset used in our study, which consists of over 39 million comments produced by over 1.8 million unique commenters, participating in discussions triggered by 320 top influencers over two countries (Brazil and Italy).

\begin{table*}[t]
\centering
    \scriptsize 
    \renewcommand{\arraystretch}{0.9}
    \caption{Dataset Overview (weekly snapshots including election dates are shown in bold in the respective country).}
    \begin{tabular}{crrrrrrrrr}%{c|rr|rr|rr|rr}
\toprule
\multirow{3}{*}{\textbf{Weekly Snapshot}} & \multicolumn{4}{c}{\textbf{Politics}}                                                  & & \multicolumn{4}{c}{\textbf{General}}                                                   \\ \cmidrule{2-10} 
                               & \multicolumn{2}{c}{\textbf{Brazil}}       & \multicolumn{2}{c}{\textbf{Italy}}        & & \multicolumn{2}{c}{\textbf{Brazil}}       & \multicolumn{2}{c}{\textbf{Italy}}        \\ \cmidrule{2-10} 
                               & \textbf{\# Posts} & \textbf{\# Commenters} & \textbf{\# Posts} & \textbf{\# Commenters} & & \textbf{\# Posts} & \textbf{\# Commenters} & \textbf{\# Posts} & \textbf{\# Commenters} \\ \midrule
1                              & 1\,487      & 37\,406      & 779           & 17\,427    &   & 746               & 172\,454                 & 733               & 54\,407        \\ %\hline
2                              & 1\,648      & 67\,799      & 739           & 20\,873    &   & 778               & 180\,711                 & 703               & 49\,290        \\ %\hline
3                              & 1\,798      & 103\,506     & 742           & 20\,876    &   & 719               & 164\,040                 & 594               & 52\,052        \\ %\hline
4                              & 1\,951      & 94\,327      & 907           & 21\,402    &   & 854               & 186\,333                 & 649               & 54\,677        \\ %\hline
5                              & \bf 2\,307  & \bf 145\,618 & 1\,080        & 22\,029    &   & 680               & 125\,414                 & 683               & 52\,318        \\ %\hline
6                              & 958         &  184\,993    & 1\,240        & 22\,890    &   & 771               & 158\,522                 & 720               & 69\,066        \\ %\hline
7                              & 1\,195      & 123\,797     & \bf 1\,316    & \bf 26\,600 &  & 723               & 131\,563                 & 657               & 61\,168        \\ %\hline
8                              & \bf 1\,400  & \bf 145\,499 &  701          & 31\,308    &   & 798               & 152\,705                 & 635               & 66\,337        \\ %\hline
9                              &  799        &  191\,282    & 762           & 17\,171    &   & 733               & 146\,128                 & 540               & 31\,520        \\ %\hline
10                             & 606         & 50\,546      & 656           & 19\,926    &   & 763               & 159\,628                 & 507               & 33\,781        \\ \bottomrule
\end{tabular}
    \label{tab:charact}
\end{table*}

\subsection{Dataset crawling}
\label{subsec:crawling}

We collected data from Instagram profiles in Brazil and Italy. Our collection targets electoral periods to capture the political debate taking place on the social network. For Brazil, we focus on Instagram posts submitted during the national general elections of October $7^{th}$ (first round) and October $28^{th}$ (second round), 2018. Our dataset covers 10 weeks (from September $2^{nd}$ until November $10^{th}$, 2018) which includes weeks before and after the election dates. Similarly, for Italy we observed the European elections held on May $26^{th}$, 2019, collecting data published from April $7^{th}$ to June $15^{th}$ (also 10 weeks). We monitor posts shared by selected profiles (see below), gathering all comments associated with those posts.

We use a custom web crawler to scrape data from Instagram that relies on the Instaloader library\footnote{\url{https://instaloader.github.io}}. We performed the crawling in September 2019. Given a profile $i$, the crawler looks for posts $i$ created during the predefined period. For each post, the crawler downloads all comments associated with it. As the interest in posts on Instagram tends to decrease sharply with time~\cite{Trevisan:2019}, we expect that our dataset includes almost all comments associated with posts created during the period of analysis. We focus only on \emph{public} Instagram profiles and posts, collecting all visible comments they received. We performed the crawling respecting Instagram rate policies to avoid overloading the service. We did not collect any sensitive information of commenters, such as display name, photos, or any other metadata, even if public. 

For each country, we monitor two groups of influencers:
\begin{itemize}
    \item \emph{Politics}: the most popular Brazilian and Italian politicians and official political profiles. We manually enumerated political leaders (e.g., congressmen, senators, governors, ministers, president) of all political parties and looked for their profiles on Instagram. We have kept only those having a large number of followers (larger than 10\,000 followers) and excluded those with minimal activity. To perform our analysis we kept the top-80 profiles, ranked by number of followers. In total, the Brazilian politics profiles created 14\,149 posts and received more than 8 million comments by
    575\,612 unique commenters during the monitored period. Similarly, the Italian profiles created 8\,922 posts, which received more than 1.9 million comments by 94\,158 distinct commenters.
    \item \emph{General}: non-political influencers used as a control group. We rely on the HypeAuditor\footnote{\url{https://hypeauditor.com/}} rank to obtain the list of most popular profiles for the Sport, Music, Show, and Cooking categories in each country. Similarly to the \textit{Politics} group, we pick $80$ profiles for each country. The Brazilian {\it general} profiles created 7\,565  posts and received 15 million comments by 295\,753 distinct commenters during the monitored period. Similarly, the Italian general profiles created 6\,421 posts and received 14 million comments carried out by 897\,421 commenters. 
\end{itemize}

\subsection{Data pre-processing}
\label{sec:preproc}

%To build the network of co-commenters, we aggregate posts by week, separately by country (Brazil and Italy) and category of influencers (general and politics). We then use the comments these posts received to build the co-commenter network, as we described in Section~\ref{sec:netw}. We build a network for each of the $10$ weeks of the datasets, country and category, obtaining $40$ networks in total. We call each a weekly-snapshot or a \emph{week} for simplicity.  

We consider only commenters who commented on more than one post when building the network for a given period $w$. This step removes 70--85\% of the commenters. We observe that 95\% of removed commenters commented less than three times when considering the complete dataset. Results presented in the following refer to the dataset after removing these occasional commenters. 

To build the network of co-commenters, we aggregate posts by considering 7-day intervals, always starting on Monday and ending on Sunday. We call each of such snapshot a \emph{week} for simplicity.  We separate our data by country (Brazil and Italy) and category of influencers (general and politics). We then use comments observed on posts created in the given internal to build the co-commenter network.

Notice that the periods covered by our data in Brazil and Italy are not coincident -- each case covers $10$ weeks in total around the respective election days. In sum, our data pre-processing generates $40$ networks, one for each of the $10$ weekly-snapshots of the 2 countries, for 2 distinct categories of influencers.

\subsection{Dataset overview}
\label{subsec:overview}

Table~\ref{tab:charact} presents an overview of our dataset, showing the numbers of posts and distinct commenters per week. Election weeks are shown in bold. In Brazil, elections were on Sunday of the $5^{th}$ and $8^{th}$ weeks ($1^{st}$ and $2^{nd}$ rounds, respectively), whereas the election in Italy took place on Sunday of the $7^{th}$ week. Focusing first on politics, we observe that the number of posts tends to steadily increase in the weeks preceding elections, reach a (local) maximum on the week(s) of the election, and drop sharply in the following. Interestingly, the largest number of commenters appears on the week immediately after the elections. Manual inspection reveals this is due to celebrations by candidates and supporters. Regarding the general category, we observe that the number of posts and commenters is rather stable, with a slight decrease in the last two weeks for Italy due to the approaching of summer holidays.

\begin{figure*}[t]
    \begin{center}
        \begin{subfigure}{0.37\textwidth}
            \includegraphics[width=\columnwidth]{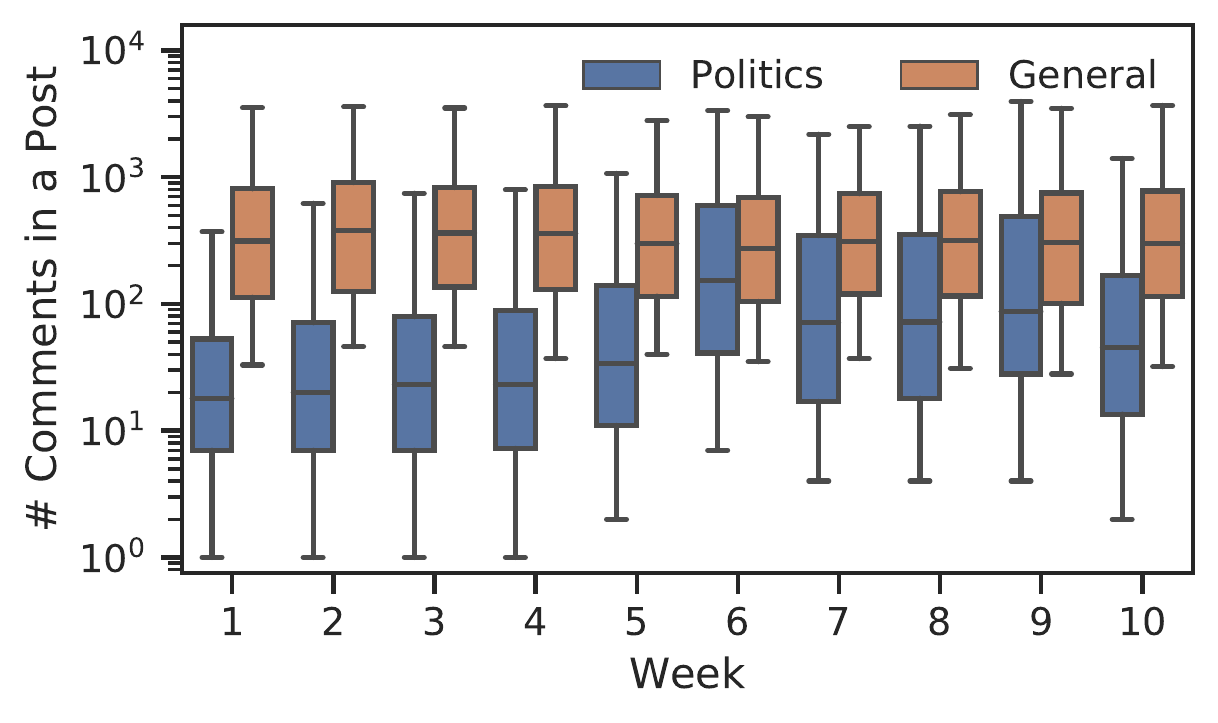}
             \caption{Brazil.}
            \label{fig:trends_br}
        \end{subfigure}
        \qquad
        \begin{subfigure}{0.37\textwidth}
            \includegraphics[width=\columnwidth]{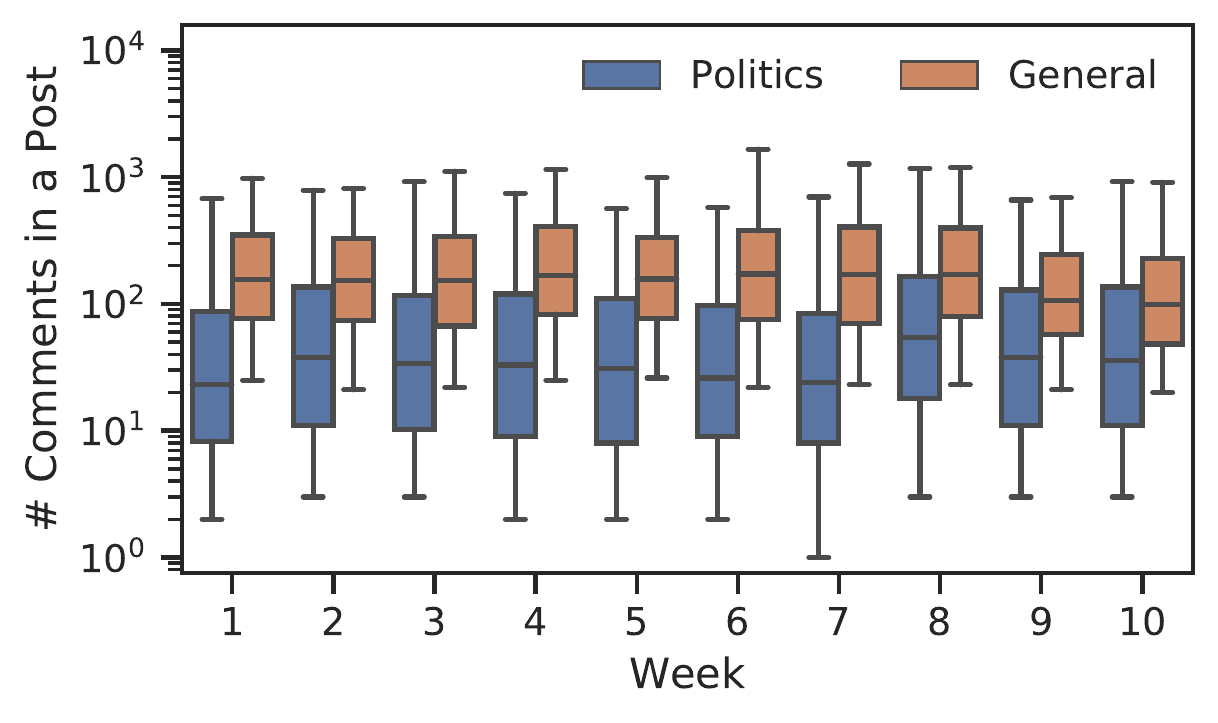}
            \caption{Italy.}
           \label{fig:trends_it}
        \end{subfigure}
        \caption{Distributions of number of comments per post (notice the log scale in $y$-axis).}
        \label{fig:trends}
    \end{center}
\end{figure*}

We complement the overview with Figure~\ref{fig:trends}, which shows the distributions of the number of comments per post during each week. We use boxplots to ease visualization. The black stroke represents the median. Boxes span from the 1\textsuperscript{st} to the 3\textsuperscript{rd} quartiles, whiskers mark the 5\textsuperscript{th} and the 95\textsuperscript{th} percentiles. For politics, the median is a few tens of comments per post, while general posts receive 10 times as much (notice the log $y$-axes). Recall that the number of distinct commenters is similar on both cases (see Table~\ref{tab:charact}), thus commenters are more active in the general profiles. Yet, posts of the main political leaders attract thousands of comments, similar to famous singers or athletes (holding for both countries). Considering time evolution, the number of comments on politics increases by an order of magnitude close to elections, with a sharper increase in Brazil.

\section{Structural analysis}
\label{sec:rq1}

We describe the network structure emerging from our data. We first illustrate characteristics of the original and  network backbones. Then, we characterize the communities and highlight insights emerging from the co-commenters backbones.

\subsection{The network backbones}
\label{sec:res_backbone}

We first show an example of network backbone, using the \nth{1} week of the Brazilian Politics scenario as case study.

\begin{figure*}[!t]
    \begin{center}
        \begin{subfigure}[t]{0.3\textwidth}
            \includegraphics[width=\columnwidth]{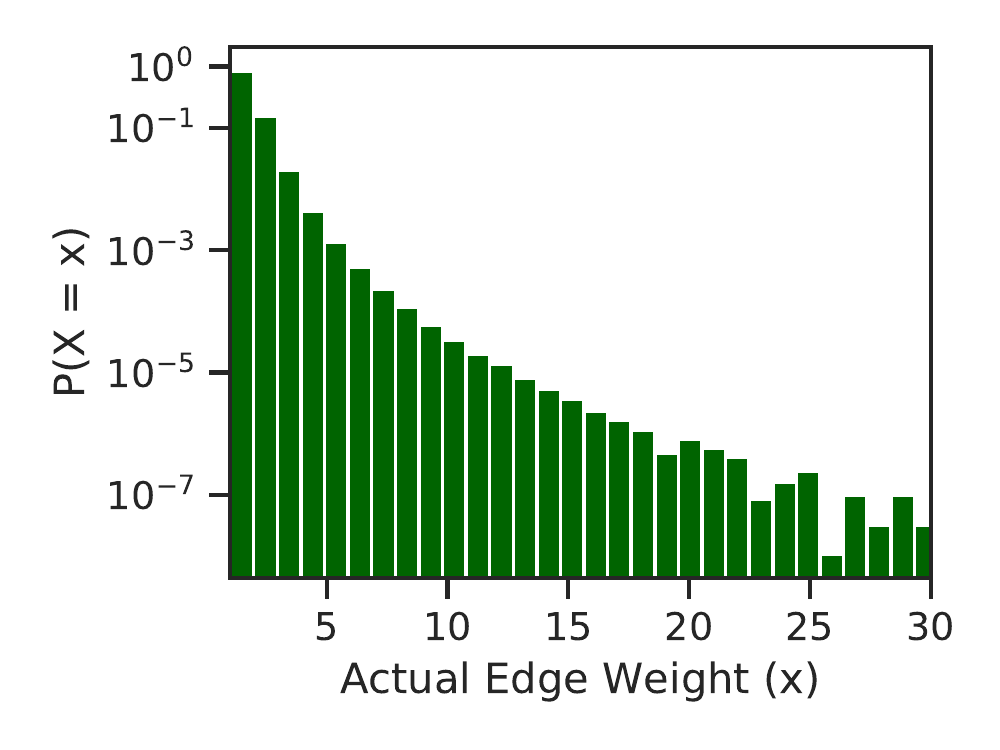}
            \caption{Probability distribution of edge weights $\gamma({cd})$ in the original graph ${G}_P$.}
            \label{fig:graph_charact_edges_backbone}
        \end{subfigure}
        \hspace{0.2cm}
        \begin{subfigure}[t]{0.3\textwidth}
           \includegraphics[width=\columnwidth]{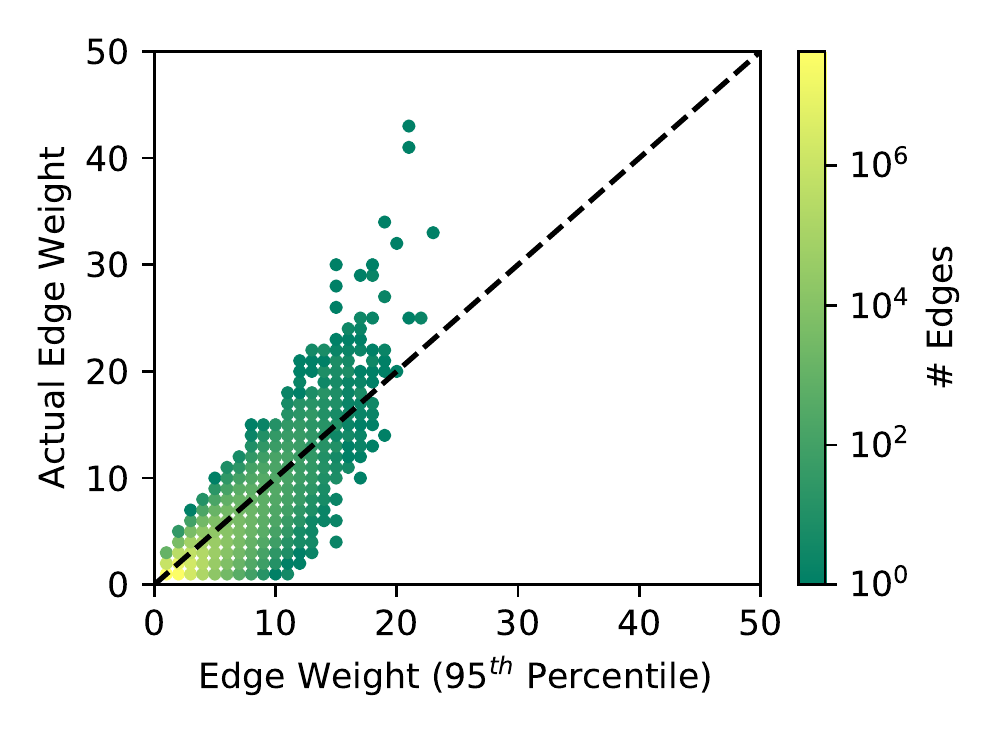}
           \caption{Actual edge weight $\gamma({cd})$ compared with null model edge weight $\hat{\gamma}_{95}({cd})$.}
            \label{fig:graph_charact_weight}
        \end{subfigure}
        \hspace{0.2cm}
        \begin{subfigure}[t]{0.3\textwidth}
          \includegraphics[width=\columnwidth]{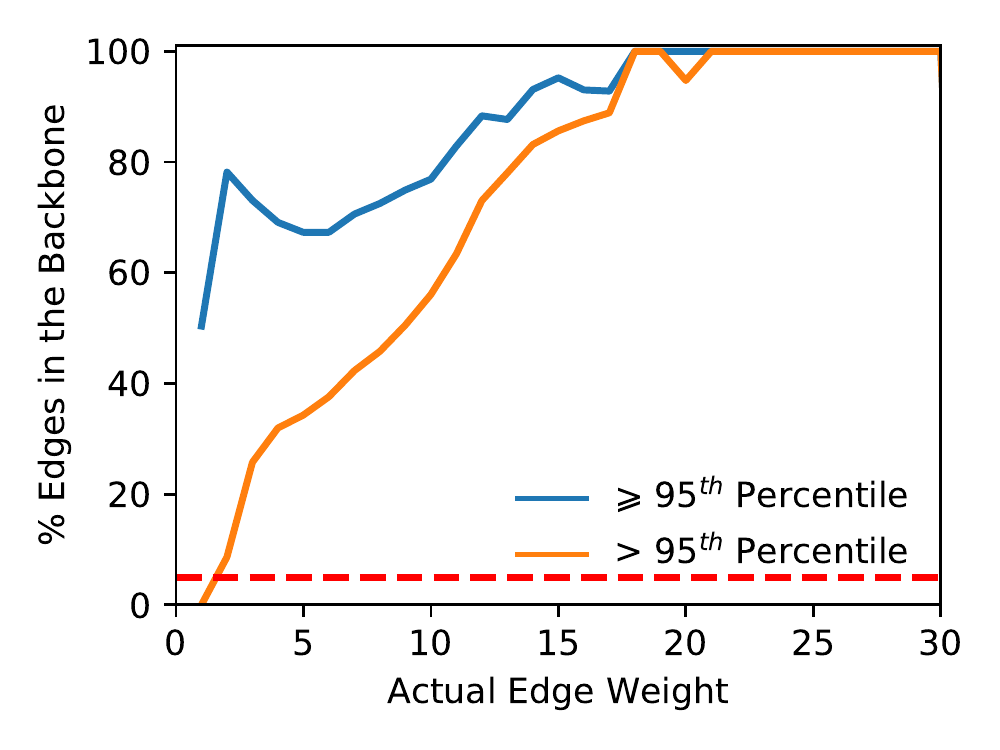}
            \caption{Fraction of edges included in the backbone ${B}_P$, separately for different values of $\gamma$.}
           \label{fig:graph_charact_actual_vs_null}
        \end{subfigure}
        \caption{Network characteristics for posts of influencers for Brazil - Politics (Week 1).}
        \label{fig:graph_charact}
    \end{center}
\end{figure*}

Figure~\ref{fig:graph_charact_edges_backbone} depicts the histogram of the edge weights in the original graph ${G}_P$. Notice that 82\% of edges have weight equal to 1, i.e., the majority of co-commenters co-comment in a single post. Higher weights are less frequent (notice the log scale on the $y$-axis). Yet, some co-commenters interact on more than 20 posts. In the following, we assess whether these weights are expected -- i.e., their weights agree with the assumption of independent user behavior.

The scatter plot in Figure~\ref{fig:graph_charact_weight} compares the observed weight in ${G}_P$ and the \nth{95} percentile of weight estimated by our reference model $\hat{G}_P$. Colors represent the number of edges, and lighter colors indicate larger quantities. Most edges have very low value for both observed and estimated weights -- notice the lightest colors for weights 1 and 2 in the bottom left corner. We are interested in the edges in which weights exceed the \nth{95} percentile of the expected weight -- i.e., those above the main diagonal. The fraction of edges over the diagonal is higher for larger weight values. This indicates that co-commenters interacting on many posts deviate from the expectation. 

Figure~\ref{fig:graph_charact_actual_vs_null} digs into that by showing the percentage of edges that are included in the network backbones separately by observed edge weight. If the null model held true, 5\% of the edges would be included (those exceeding the \nth{95} percentile) -- highlighted by the red dotted line. But in ${G}_P$, edges weights do not always follow the null hypothesis of independent behavior, especially for edges with large weights.

It is also important to remark that ${G}_P$ edge weights are integer numbers, and our generative model provides discrete distributions. Therefore, the computation of percentiles is critical since the same value can refer to a \emph{range} of percentiles. This causes a rounding issue that is critical for low values. Filtering weights \emph{greater than} or \emph{greater or equal to} particular values results in significant differences for low weights. Figure~\ref{fig:graph_charact}c illustrates it by reporting the fraction of edges that would be included in the backbone in the two cases. Using \emph{greater than} corresponds to a conservative choice since we include only edges for which the expected weight is strictly higher than the \nth{95} percentile (orange curve). Notice how the number of edges in the backbone is reduced for low weights. Conversely, \emph{greater or equal to} would preserve more edges, including those whose weight possibly corresponds to a lower percentile (blue curve). We here maintain a \emph{conservative} choice and keep edges whose actual weight is strictly greater than the \nth{95} percentile.

\begin{table}[t]
    \footnotesize
    \centering
    \caption{Characteristics of the original network ${G}_P$ and  network backbone ${B}_P$ for Brazil - Politics (Week 1).}
    \label{tab:backbone_example}
    \begin{tabular}{|c|c|c|c|c|}
        \toprule
        \textbf{Network} & \textbf{\# Nodes} & \textbf{\# Edges} & \textbf{\# Comm} & \textbf{Modularity} \\ \midrule
        Original         & 37 k            & 74.09 M          & 6                & 0.22                \\ %\hline
        Backbone         & 26 k (70.7\%)            & 1.06 M   (1.4\%)      & 19               & 0.59                \\ \bottomrule
    \end{tabular}
\end{table}

Table~\ref{tab:backbone_example} describes the resulting  network backbone ${B}_P$ after filtering, comparing it with the original graph ${G}_P$. We focus on week 1 here, but results are consistent for all weeks. Our approach discards 98.6 \% of the edges -- i.e., the vast majority of them is not salient. We remove 29\% of nodes, which remain isolated in ${B}_P$. To highlight the benefits of the approach, we include the number of communities and the modularity in the original and backbone graphs. The Louvain algorithm identifies only 6 communities with very low modularity in the original graph. On the backbone, it identifies more communities, and modularity increases from $0.22$ to $0.59$. 

Table~\ref{tab:backbone_time} summarizes the main characteristics of the network backbones obtained on each week for Brazil, Politics. Focusing on the first four columns, notice that we still include the majority of nodes, with percentages ranging from 68\% to 95\%. Considering edges, the percentage is always low (0.6--2.6\%). The fourth column reports the fraction on edges in the backbone having weight larger than 1. Remind that, by design, a random behavior would lead to 5\% of edges in the backbone, while here we observe up to 19\%, despite our conservative filtering criteria. Results are rather stable and consistent over time.

\subsection{Communities of commenters}
\label{sec:res_communities}

We now study the communities obtained from the  backbone graphs. The last two columns of Table~\ref{tab:backbone_time} show that we obtain from $19$ to $32$ communities, depending on the week. Modularity values are high (always above $0.5$), meaning that the community structure is strong.

\begin{table}[t]
    \footnotesize
    \centering
    \caption{Breakdown of backbone and communities over different weeks for Brazil, Politics. In bold, the weeks of the elections.}
    \label{tab:backbone_time}
    \begin{tabular}{|c|r|r|r||r|r|}
    \toprule
    \textbf{Week} & \textbf{\% Nodes} & \textbf{\% Edges} & \makecell{ \textbf{\% Edges} \\ ${\gamma}{(cd)}>1$  } & \textbf{\# Comm} & \textbf{Mod.} \\ \midrule
    1             & 70.69             & 1.40                       & 11.43                                     & 19               & 0.59                \\ 
    2             & 93.36             & 2.11                       & 12.19                                     & 27               & 0.64                \\ 
    3             & 73.81             & 1.01                       & 4.75                                      & 20               & 0.52                \\ 
    4             & 93.63             & 2.23                       & 15.10                                     & 32               & 0.69                \\ 
   \bf 5             & \bf 94.30             &\bf 2.65                       & \bf 19.36                                     &  \bf 17               & \bf 0.61                \\ 
    6             & 91.49             & 2.36                       & 19.37                                     & 31               & 0.66                \\ 
    7             & 94.05             & 1.87                       & 15.45                                     & 31               & 0.66                \\ 
    \bf 8             & \bf 95.40             & \bf 2.13                       &  \bf15.29                                     &  \bf 27               & \bf 0.64                \\ 
    9             & 68.01             & 0.62                       & 4.06                                      & 24               & 0.59                \\ 
    10            & 71.33             & 1.11                       & 7.21                                      & 29               & 0.61                \\ \bottomrule
    \end{tabular}
\end{table}

We summarize results for the other scenarios in Table~\ref{tab:sum_backbone_networks}, reporting only average values across the 10 weeks. First, focusing on Politics and comparing Brazil and Italy (first two rows), we observe similar percentages of nodes in the network backbones. For Italy a larger fraction of edges are retained, potentially because of the smaller volume of profiles and comments (see Section~\ref{sec:dataset}). For Brazil, we obtain a larger number of communities with higher values of modularity than in Italy. 

\begin{table}[b]
    \centering
    \footnotesize
    \caption{Networks backbone and identified communities for Brazil (BR) and Italy (IT). We show average values over the 10 weeks.}
    \label{tab:sum_backbone_networks}
    \begin{tabular}{|c|c|c|c||c|c|}
    \toprule
    \textbf{Scenario} & \textbf{\% Nodes} & \textbf{\% Edges} & \makecell{ \textbf{\% Edges} \\ ${\gamma}{(cd)}>1$  } & \textbf{\# Comm} & \textbf{Mod.} \\ \midrule
    BR Politics   & 84.61                 & 1.81                   & 12.42                            & 26               & 0.62              \\ 
    IT Politics    & 87.33                  & 3.39                   & 21.79                            & 11               & 0.44              \\ \midrule 
    BR General    & 65.35                  & 0.82                   & 8.83                             & 81               & 0.79              \\ 
    IT General     & 60.03                  & 2.23                   & 12.57                            & 48               & 0.72              \\ \bottomrule
    \end{tabular}
\end{table}

Moving to the General scenarios (\nth{3} and \nth{4} rows), we notice that fewer nodes and edges are in the backbones compared to Politics. Interestingly, we identify more and stronger communities. We root this phenomenon in the heterogeneity of the General scenarios that include influencers with different focuses, potentially attracting commenters with different interests. Manual inspection confirms the intuition -- i.e., we find some communities interested in sports, others on music, etc. For politics, instead, we find a more tangled scenario. Even if communities are rather strong, some of them include profiles commenting on politicians of different parties and embracing different topics. Next, we evaluate communities in the Politics scenario.

\subsection{Analysis of political communities}
\label{sec:res_insights}

We now focus on Politics and show how the activity of commenters spreads across political profiles of different parties. Here we focus on the election week for both countries to better capture the peak of the political debate on Instagram.

\begin{figure}[!t]
     \begin{center}
        \begin{subfigure}[t]{\columnwidth}
        \centering
            \includegraphics[width=0.95\columnwidth]{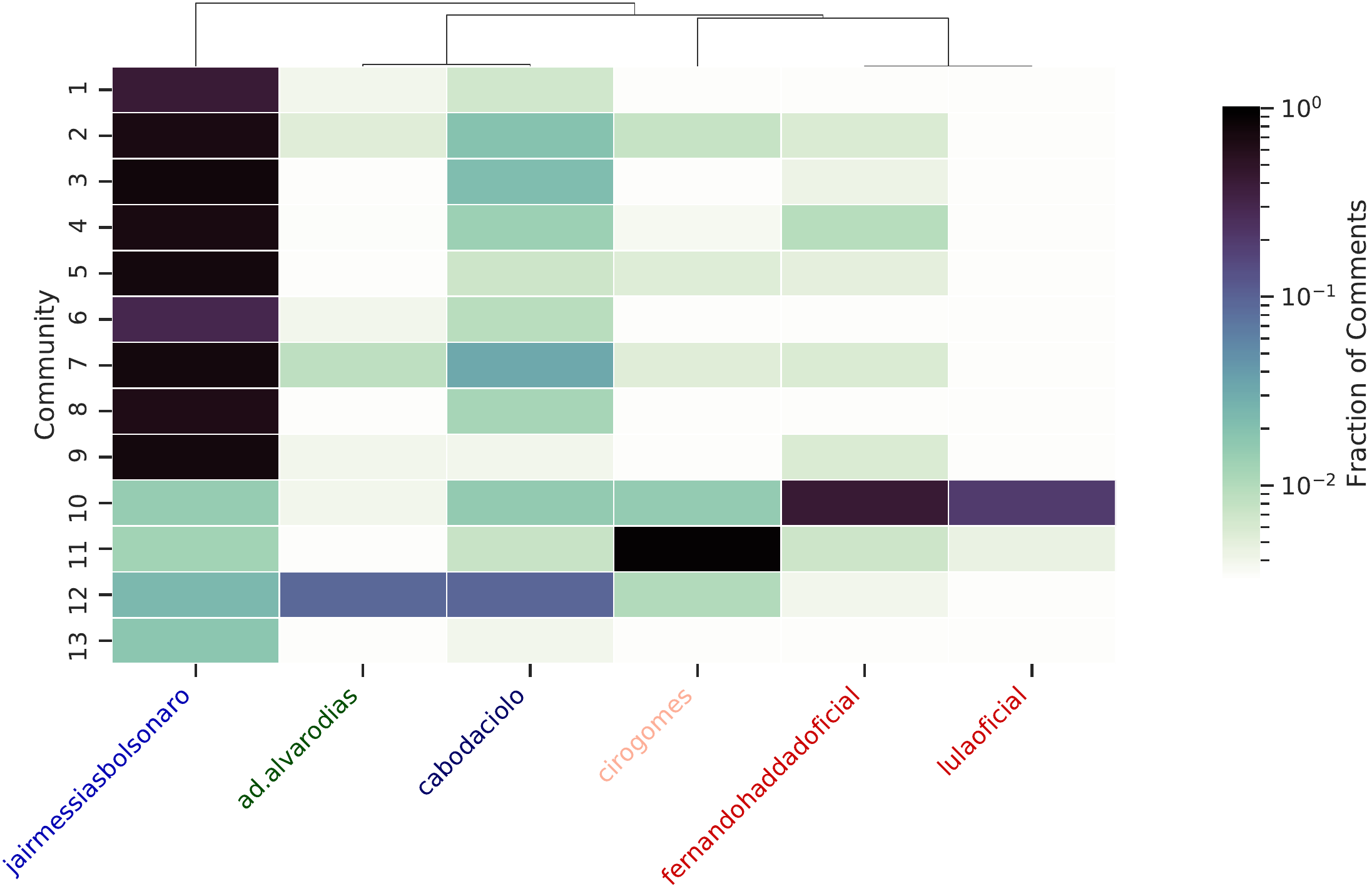}
            \caption{Brazil (1st round)}
            \label{fig:hetmaps_br}
        \end{subfigure}
        \begin{subfigure}[t]{\columnwidth}
        \centering
        \vspace{5mm}
           \includegraphics[width=0.99\columnwidth]{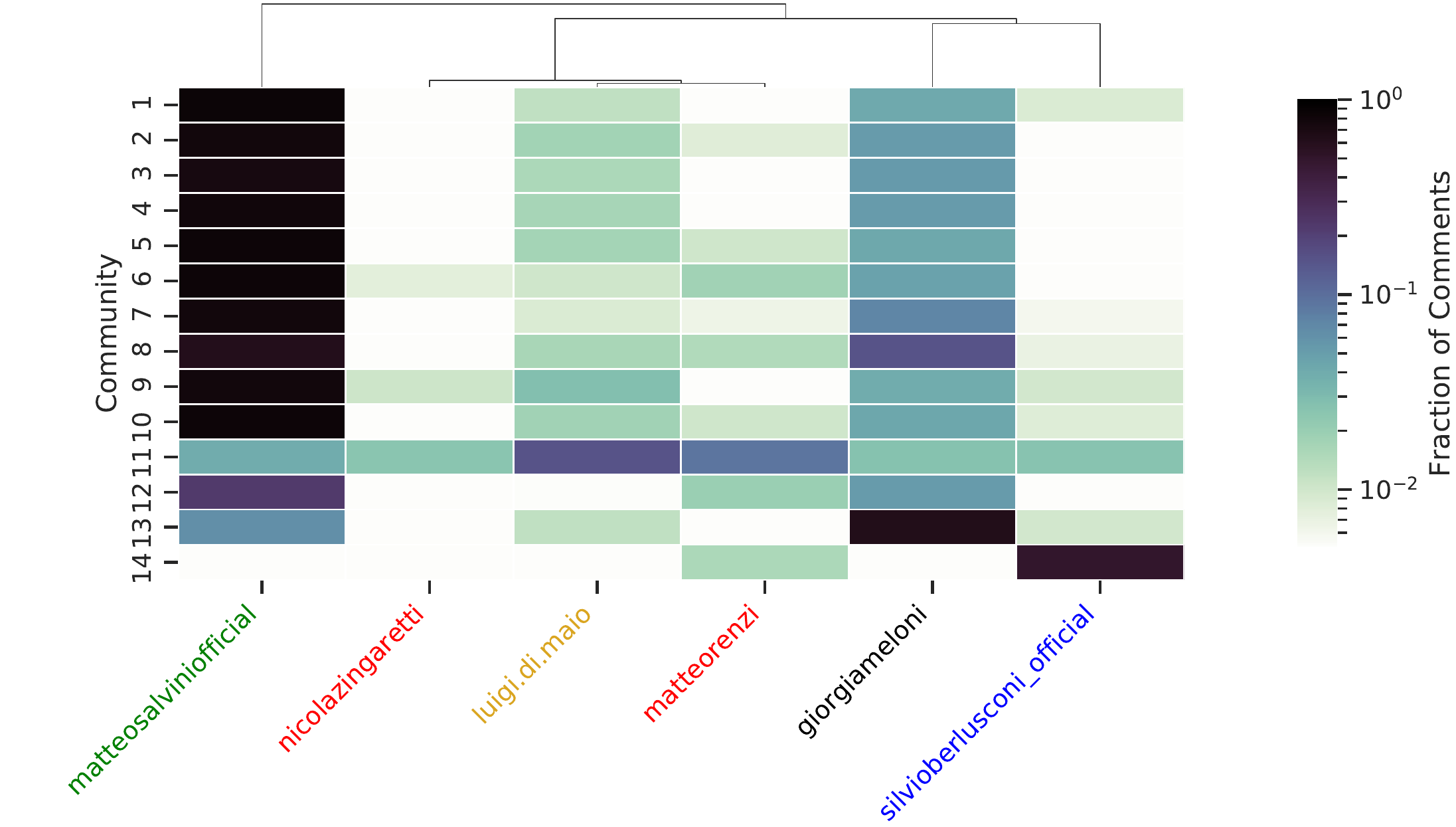}
           \caption{Italy}
            \label{fig:hetmaps_it}
        \end{subfigure}
        \caption{Distribution of comments among political leaders for each community during the main election weeks.}
        \label{fig:hetmaps}
     \end{center}
\end{figure}

We first focus on the main political leaders of the two countries and study how the communities of co-commenters distribute their interests among their posts. We consider six politicians in each country. Figure~\ref{fig:hetmaps} shows how the commenters of each community are spread among posts of each politician using a heatmap. Columns represent politicians and rows represent communities. The color of each cell reflects the fraction of the comments of the community members that are published on the posts of the politician.

\begin{figure*}[!h]
    \begin{center}
        \begin{subfigure}{0.85\textwidth}
            \includegraphics[width=\columnwidth]{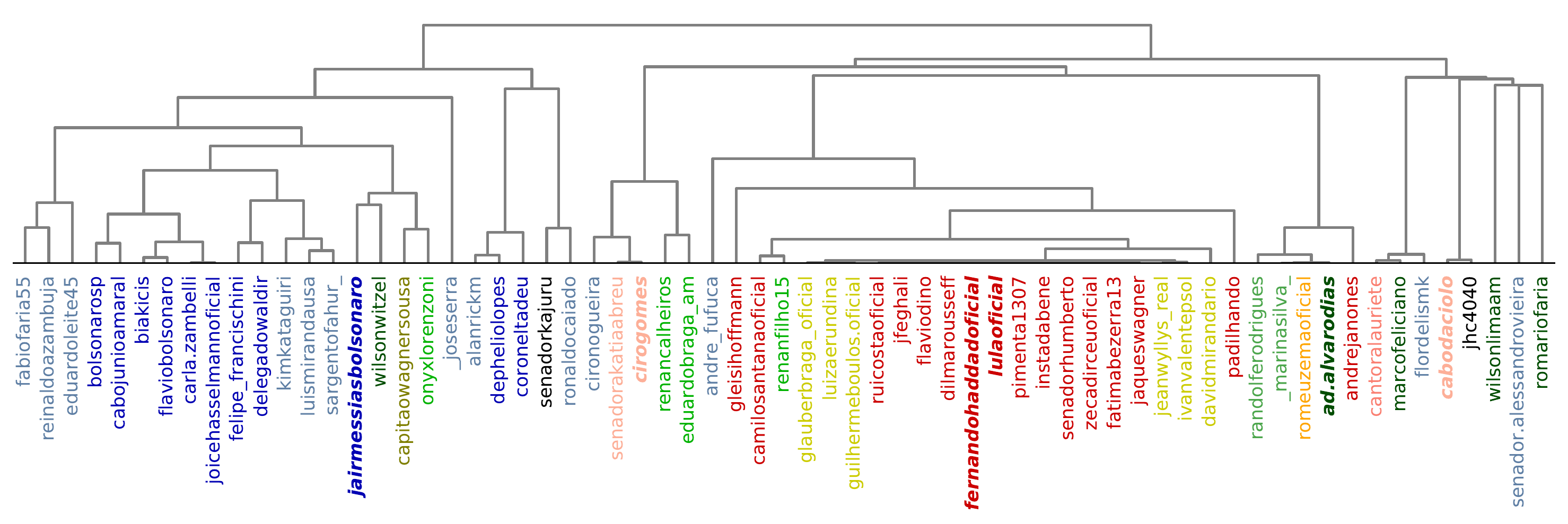}
            \caption{Brazil.}
           \label{fig:dendo_br}
        \end{subfigure}
        \begin{subfigure}{0.85\textwidth}
            \includegraphics[width=\columnwidth]{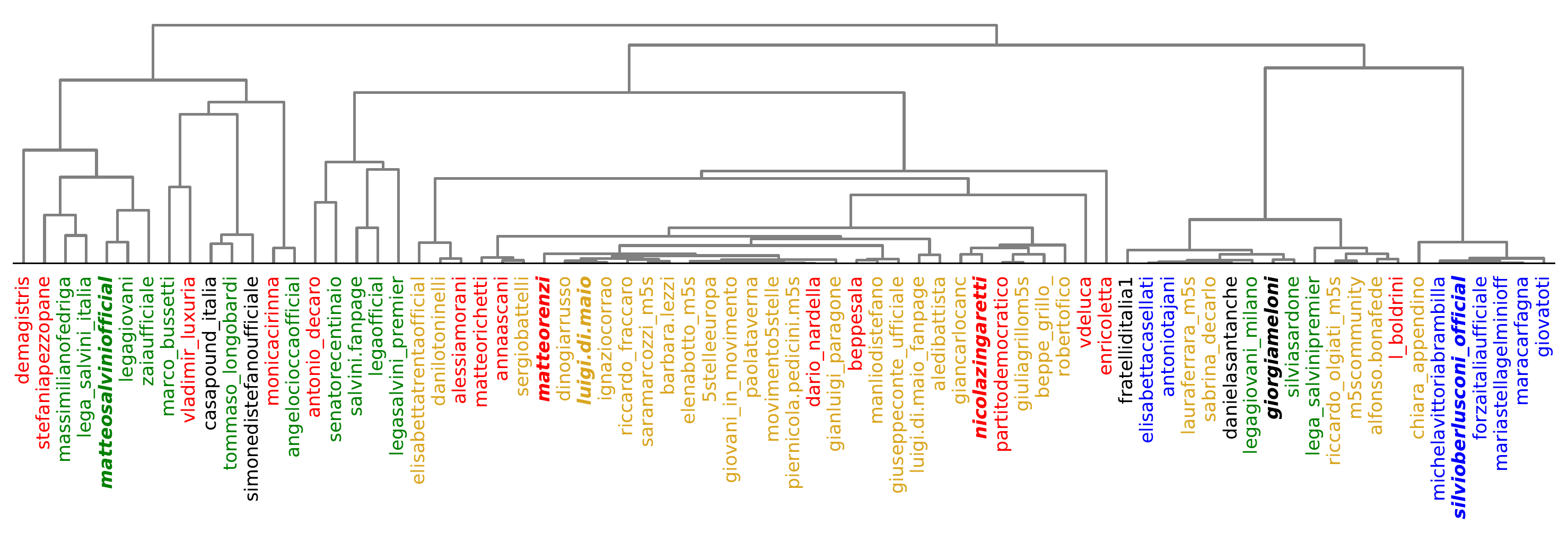}
            \caption{Italy.}
           \label{fig:dendo_it}
        \end{subfigure}
        \caption{Dendogram of political influencers clustered according to commenter communities. Influencers are colored according to their political coalition. }
        \label{fig:dendo}
    \end{center}
\end{figure*}

To gauge similarity of profiles, the top of the heatmaps report a dendrogram that clusters politicians based on the communities of their commenters. We define as similarity metric of politicians the Pearson correlation among the activity of communities on their posts. In other words, we compare them by computing the correlation between the corresponding columns of the heatmap. Hence, two politicians that receive comments from the same communities have high similarity.

Looking at the Brazilian case (Figure~\ref{fig:hetmaps_br}), we notice that most communities are interested in a single candidate - Jair Bolsonaro (jairmessiasbolsonaro), with the large majority of comments focused on his posts. This behavior is expected given his large number of followers and popularity. Indeed, communities $1-9$ comment almost uniquely on Bolsonaro.  Focusing on the dendrogram on the top of the figure, Bolsonaro has the highest dissimilarity from the others, i.e., he is the first candidate to be separated from others. Other clusters reflect accurately the candidates' political orientation. Left-leaning candidates (Ciro Gomes, Fernando Hadaad and Luiz Inacio Lula\footnote{Haddad replaced Lula, who was barred by the Brazilian Election Justice.}) are close, as well as the ones leaning towards the right-wing parties (Alvaro Dias, Cabo Daciolo and Jair Bolsonaro). 

Similar considerations hold for the Italian case (Figure~\ref{fig:hetmaps_it}). Communities $1-10$ focus on Matteo Salvini (matteosalviniofficial). He is the only one for which we identify multiple and well-separated communities. The other right-wing leaders have communities active almost exclusively on their posts, e.g., communities $13$ and $14$ for Silvio Berlusconi and Giorgia Meloni. Other leaders (e.g., Matteo Renzi and Nicola Zingaretti for the Democratic Party and Luigi Di Maio for the Five Star Movement) share a large fraction of commenters in community $11$. This suggests these commenters are almost equally interested in the three leaders. Indeed, looking at the dendrogram, these last three profiles are close to each other. Matteo Salvini (leader of the most popular party) has the maximum distance from others. Similar to the Bolsonaro's case, Salvini is a single leader who polarizes communities, thus well-separated from others.

We now broaden the analysis to all politicians. We label each politician according to his/her political \textit{coalition} using available public information.\footnote{Differently from e.g., the US or UK, in both Brazil and Italy the political system is fragmented into several parties that form coalitions during and after elections~\cite{Ferreira:2019}.} For Brazil, we rely on the Brazilian Superior Electoral Court,\footnote{\url{http://divulgacandcontas.tse.jus.br/divulga/\#/estados/2018/2022802018/BR/candidatos}} while for Italy we use the official website of each party. Rather than reporting the activity of each community on all politicians, we show only the dendrograms that cluster them, following the same methodology used in Figure~\ref{fig:hetmaps}.

Figure~\ref{fig:dendo} shows the results, where the party leaders/candidates shown in Figure~\ref{fig:hetmaps} are marked in bold. Politicians of the same parties appear close, meaning that their posts are commented by the same communities. For Brazil, the higher splits of the dendrogram roughly create two clusters, for left and right-wing parties. In Italy, we can identify three top clusters, reflecting the tri-polar system. 
Less expected are the cases in which politicians from distant political leanings attract the interest of the same communities and are close in the dendrogram. For example, in Italy, we find the profile of Monica Cirinn\`{a} (left-wing) very close to Angelo Ciocca (right-wing). Manual inspection reveals a considerable number of disapproving comments to posts of the first politician that are published by commenters supporting the second. The same happens for Vladimir Luxuria, whose some supporters disapprove Marco Bussetti's posts (and vice-versa). The structure of the backbone graph reflects the presence of profiles that bridge communities.

In sum, our methodology uncovers the structure of communities, which reflect people's engagement to politicians over the spectrum of political orientation. Most communities are well-shaped around single profiles, but sub-communities emerge too, possibly around particular topics, as we will check next. In some cases, commenters cross the expected political divide, commenting on profiles from different political orientations.

\section{Textual properties of discussions}\label{sec:rq2}

We now focus on how the communities differ in terms of textual, sentiment and psychological properties of comments. 

\subsection{Political communities' interests}

We now look into how communities in politics are attracted by different posts. Since communities differ in the number of members and in the number of comments they post, we consider a relative \emph{activity index} of the community in a post, given by the fraction of the community's comments going to the post. We use data from the week of the main elections in Brazil (week 5).

\begin{figure*}[!t]
    \begin{center}
        \begin{subfigure}{0.35\textwidth}
        \centering
            \includegraphics[width=0.95\columnwidth]{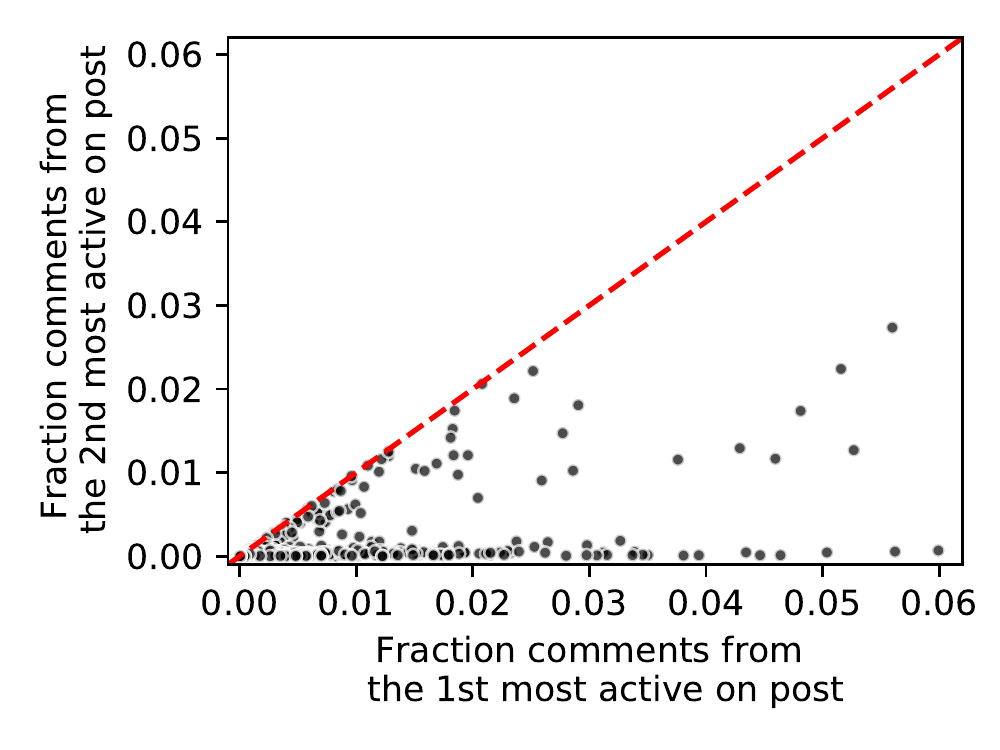}
            \caption{Index of activity of the 1$^\textrm{st}$ and $2^\textrm{nd}$ most active communities. Each point represents a post.}
          \label{fig:Activity1}
        \end{subfigure}
        \hspace{0.3cm}
        \begin{subfigure}{0.35\textwidth}
        \centering
        \vspace{0.15cm}
             \includegraphics[width=\columnwidth]{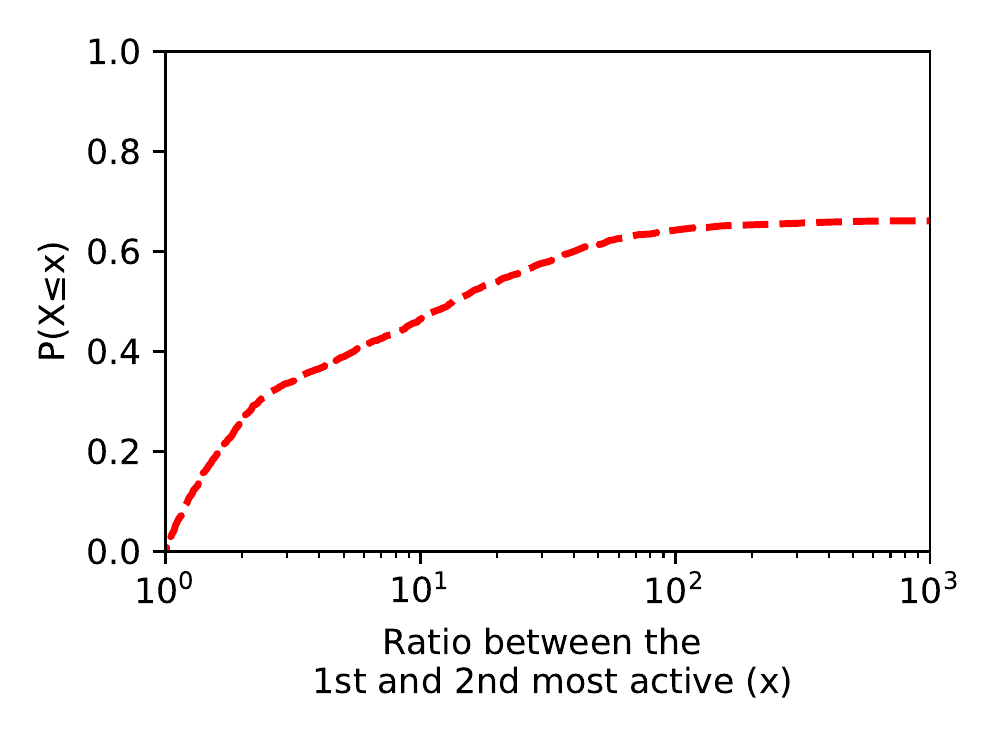}
            \caption{Ratio between the index of the 1$^\textrm{st}$ and 2$^\textrm{nd}$ most active community on posts.}
          \label{fig:Activity2}
        \end{subfigure}
        \caption{Activity of communities on posts.}
        \label{fig:Activity}
    \end{center}
\end{figure*}

Figure~\ref{fig:Activity}a quantifies, for each post, the two most active communities. The $x$-axis reports the index for the community with the highest activity on the post, while the $y$-axis reports the index for the second most active community in the post. We observe that, in all cases, the most active community leaves less than 7\% of its comments in a unique post (see the $x$-axis). Given there are $2\,144$ posts in this snapshot, even a relative activity of 1\% could be considered highly concentrated attention, suggesting that communities are built around specific posts. In  $\approx 40$\% of the posts, the relative activity of the second most active community ($y$-axis) is very low compared to the most active one. We quantify this in Figure~\ref{fig:Activity}b, which reports the ratio between the relative activity of the first and the second most active communities. We observe that, in the 55\% of cases, the most active community has at least 10 times higher index than the second one -- notice the x-axis log-scale. Hence, we have strong shreds of evidence that communities are attracted by specific posts.

\begin{figure*}[!t]
    \begin{center}
        \begin{subfigure}{0.20\textwidth}
        \centering
            \includegraphics[width=\columnwidth]{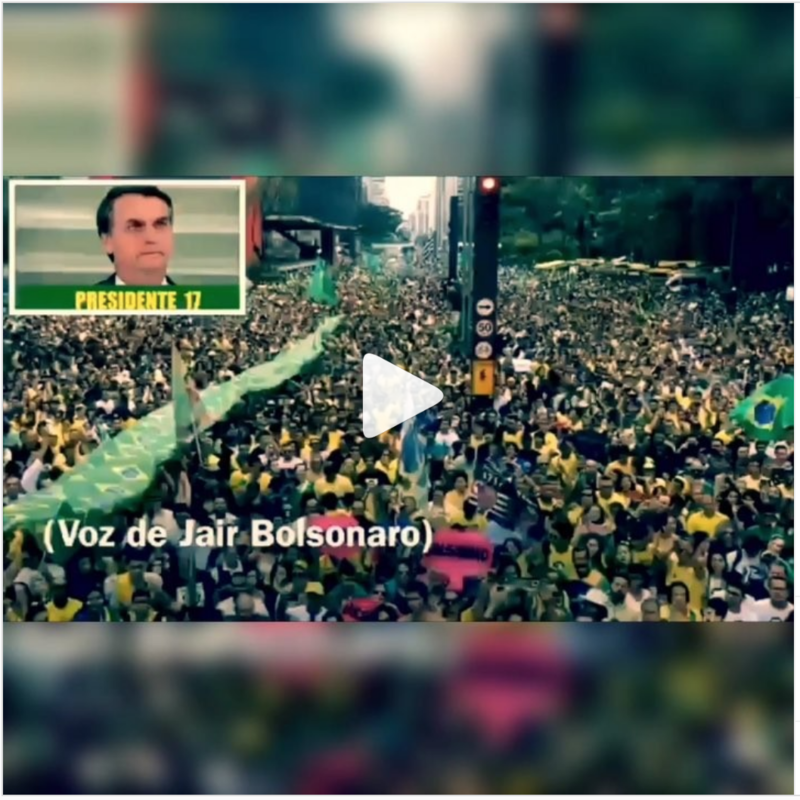}
            \caption{\href{https://www.instagram.com/p/BoXpvV6Hrkk}{Community 3 - Post about a rally in São Paulo.\\ \url{www.instagram.com/p/BoXpvV6Hrkk}}}
          \label{fig:manifest_1}
        \end{subfigure}
        \hspace{0.3cm}
        \begin{subfigure}{0.20\textwidth}
        \centering
            \includegraphics[width=\columnwidth]{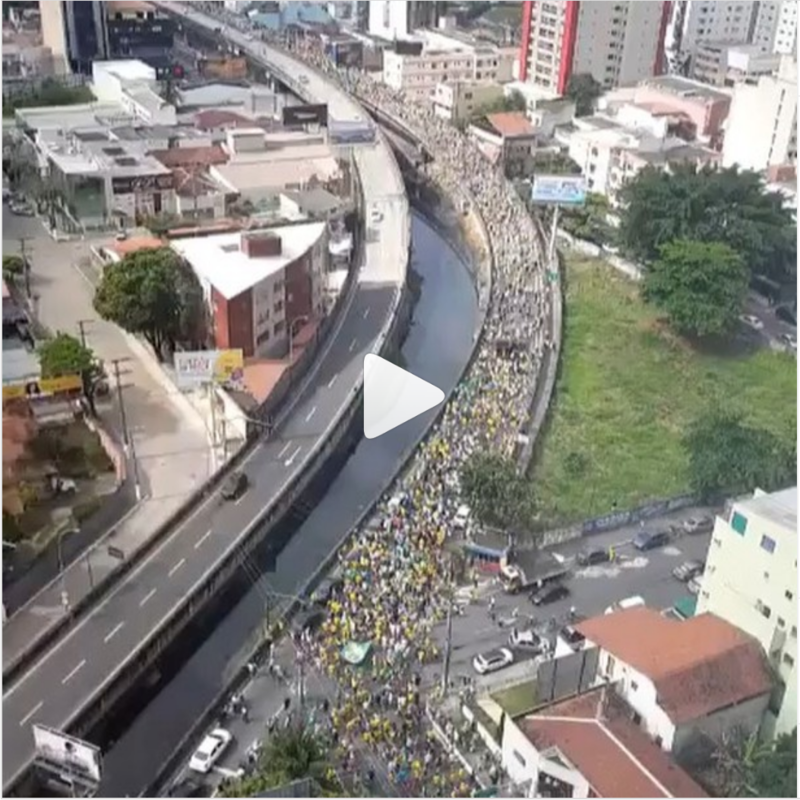}
            \caption{\href{https://www.instagram.com/p/BoXMwvwn6xj}{Community 3 - Post about a rally in Vitória.\\ \url{www.instagram.com/p/BoXMwvwn6xj}}}
          \label{fig:manifest_2}
        \end{subfigure}
        \hspace{0.3cm}
        \begin{subfigure}{0.20\textwidth}
        \centering
            \includegraphics[width=\columnwidth]{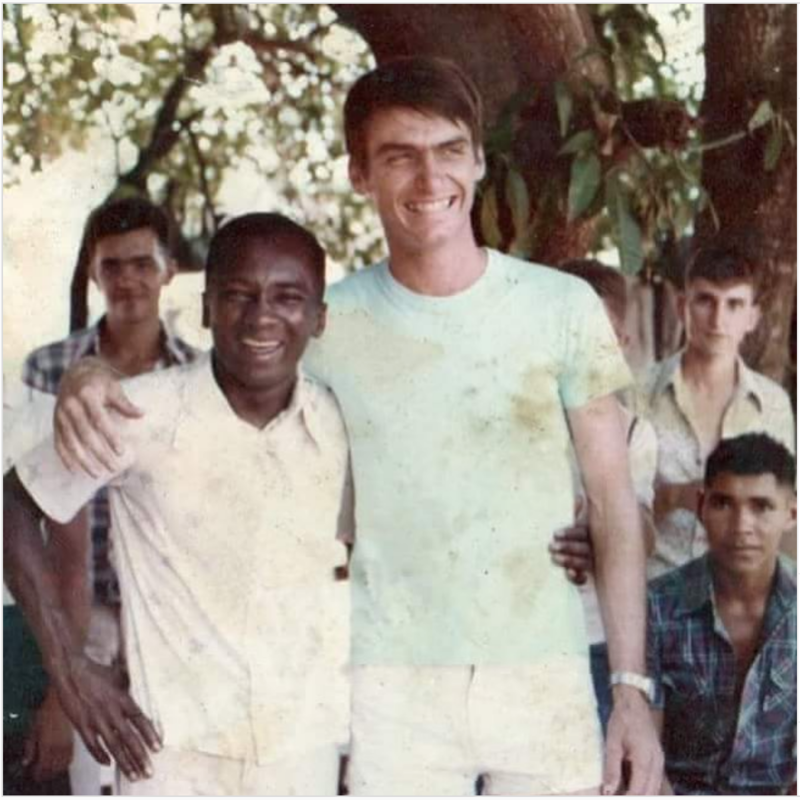}
            \caption{\href{https://www.instagram.com/p/BomRItfH9p8}{Community 7 - Post discussing racism.\\ \url{www.instagram.com/p/BomRItfH9p8}}}
        \label{fig:racism_1}
        \end{subfigure}
        \hspace{0.3cm}
        \begin{subfigure}{0.20\textwidth}
        \centering
            \includegraphics[width=\columnwidth]{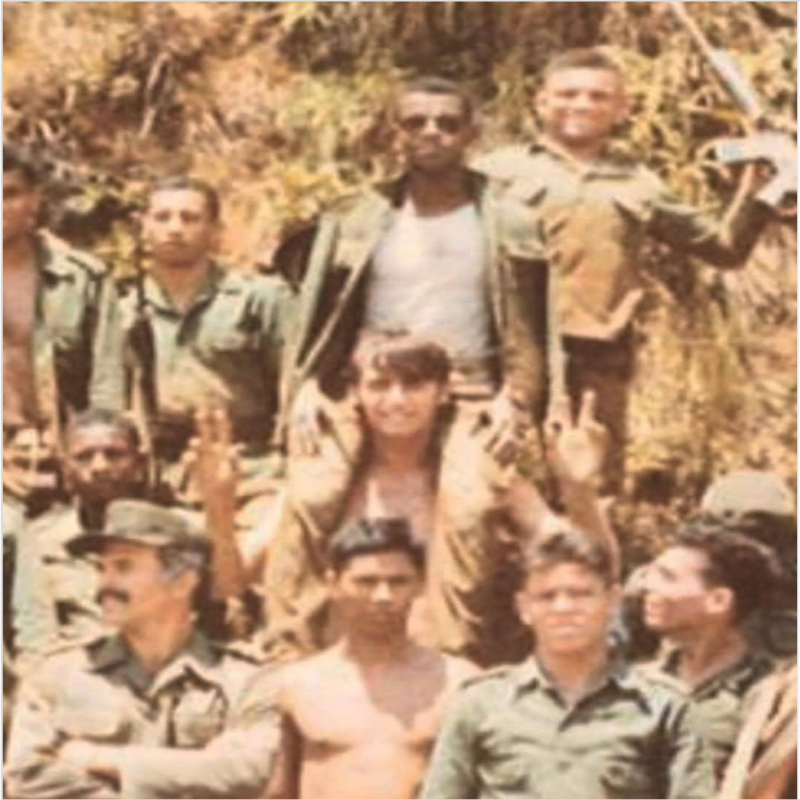}
            \caption{\href{https://www.instagram.com/p/Boe7fQcHfJB}{Community 7 - Another post discussing racism. \\ \url{www.instagram.com/p/Boe7fQcHfJB}}}
          \label{fig:racism_2}
        \end{subfigure}
        \caption{Examples of posts by Jair Bolsonaro (jairmessiasbolsonaro) in which two communities show high interest.}
        \label{fig:examples}
    \end{center}
\end{figure*}

Figure~\ref{fig:examples} shows posts that attracted high interest from communities $3$ and $7$, which we use as running examples along with communities $10$ and $11$. Community $3$ comments mostly on posts related to public events Bolsonaro promoted via Instagram (as in Figures~\ref{fig:examples}a and Figures~\ref{fig:examples}b), while community $7$ comments on posts where the candidate tries to show his proximity with black people to debunk his associations with racism (Figures~\ref{fig:examples}c and Figures~\ref{fig:examples}d).

\subsection{Properties of communities' comments}

We now take all communities found in each week and extract properties of the comments of their members, namely: i) Average comment length (in characters); ii) Fraction of comments that include at least one mention; iii) Average number of hashtags per comment; iv) Fraction of comments with at least one uppercase word; v) Average number of comments per commenter; vi) Average number of emojis per comment; and vii) Fraction of replies among comments. Together these metrics  capture important aspects of the communities' behavior. For example, the comment length, the number of emojis per comment and the use of uppercase words (commonly associated with a high tone) can describe the way the communities interact on Instagram. Mentions, the use of hashtags and replies are strongly associated with engagement, information spreading and direct interaction of commenters, respectively.

\begin{figure*}[t]
    \centering
    \subfloat[PCA]{\includegraphics[width=0.8\columnwidth]{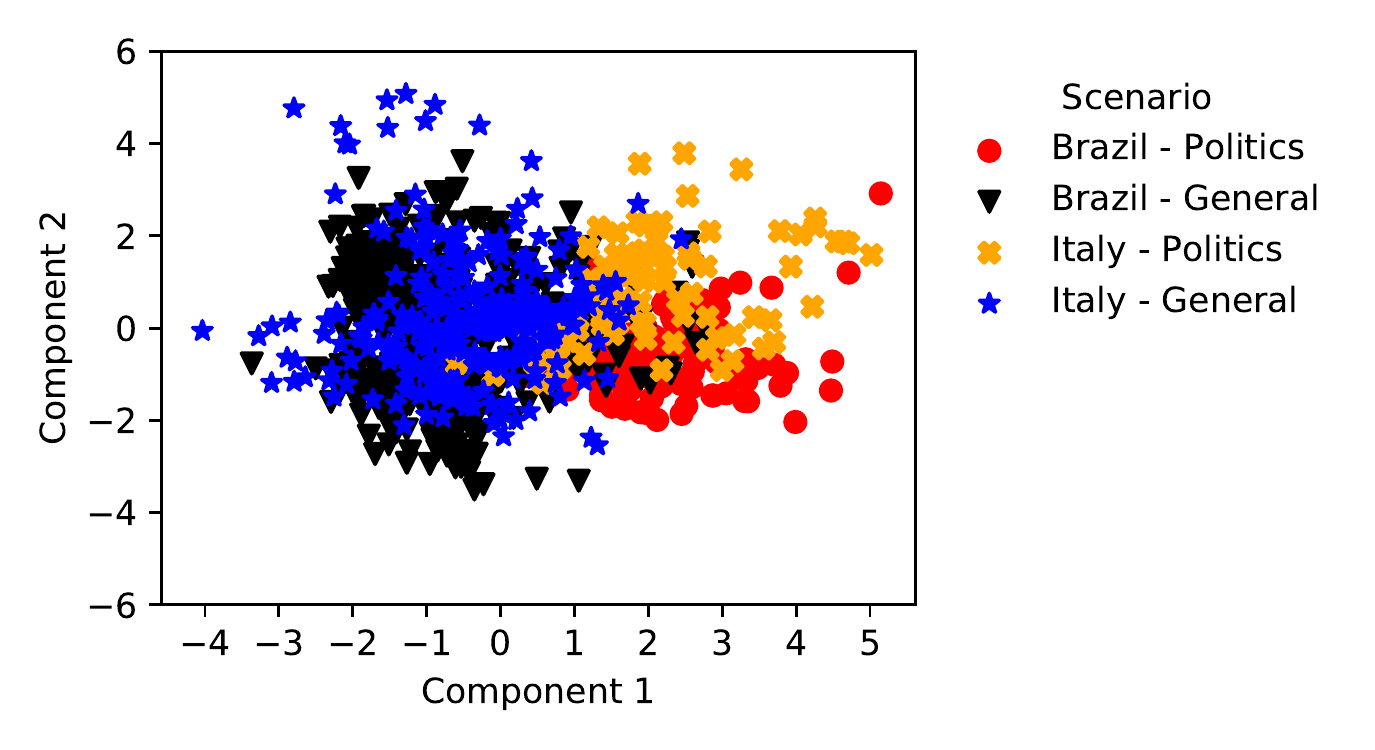}\label{Fig:PCA}}
    \subfloat[Description of the 2 first principal components]{\includegraphics[width=0.85\columnwidth]{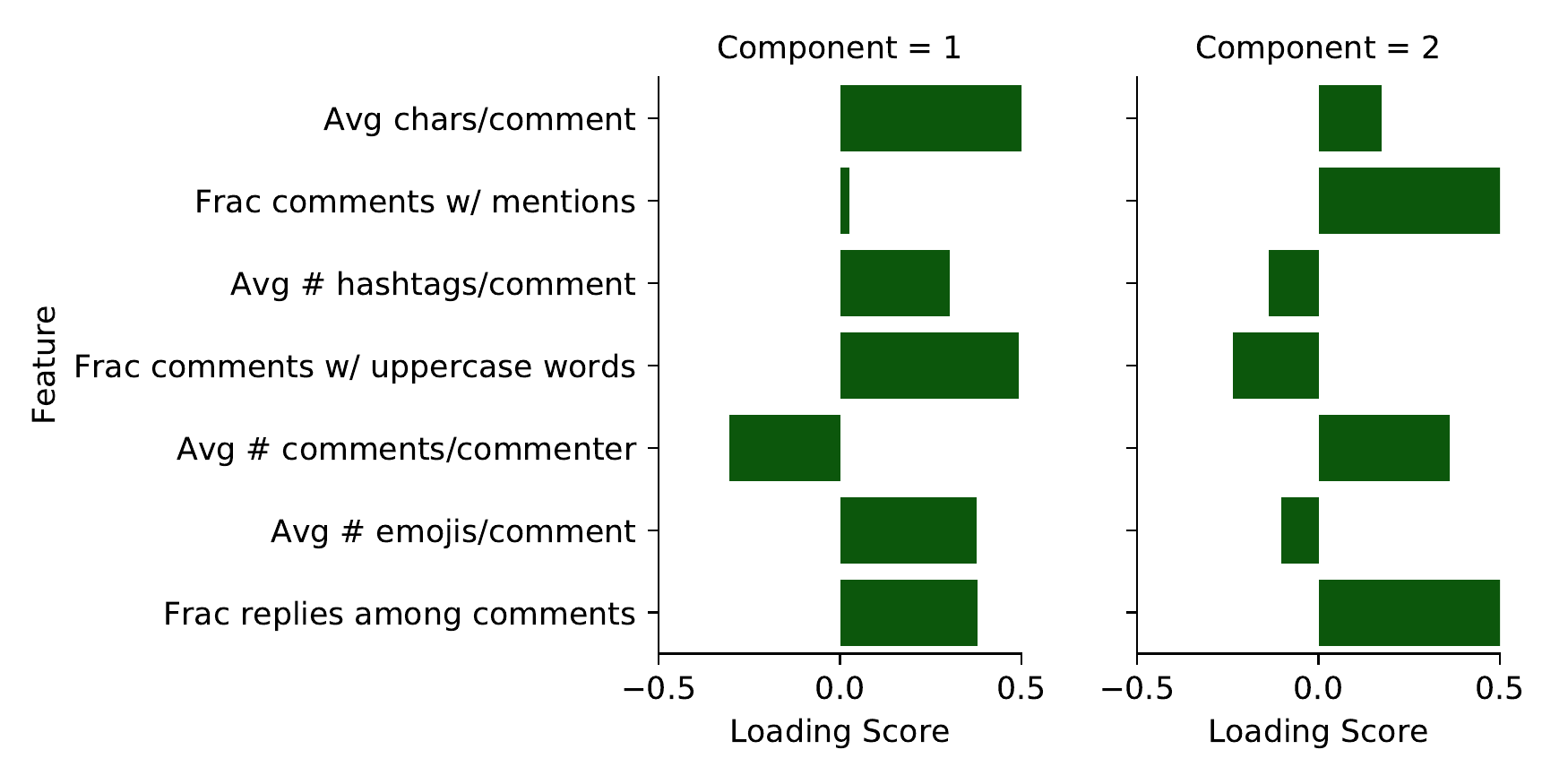}\label{Fig:Explain}}

    \caption{(a) 2-D representation of communities based on seven metrics using PCA. (b) Description of the two principal components in terms of the original metrics; the bar represents the loading scores for the components (positive or negative).}
\end{figure*}

We study the communities by applying Principal Component Analysis (PCA) to vectors that represent communities using the seven previously described metrics. PCA is a well-known method for dimensionality reduction in multivariate analysis. It projects the data along its principal components (PCs), i.e., axes that capture most of the variance in the data~\cite{Tipping:1999}. Figure~\ref{Fig:PCA} shows the representation obtained for each community using the two principal components, where the color represents the pair country-scenario. The 2-D representations of communities for both politics scenarios are more tightly clustered and overlapping than for the general scenario. This behavior suggests that, when considering the given features, communities on politics are more homogeneous than the communities on the general scenario.

To understand which metrics best distinguish the communities in Figure~\ref{Fig:PCA}, we study the \emph{loading scores} for the two principal components. The loading score quantifies the contribution of each metric to a principal component. The largest the score (in absolute value) the more the metric contributes to the component (positively or negatively).

In Figure~\ref{Fig:Explain} bars represent the magnitude of loading scores for each metric for the PC 1 (left) and PC 2 (right). The PC 1 (left) can be associated with lengthy comments, high usage of uppercase, emojis, replies and hashtags, and a low number of comments per commenter. From Figure~\ref{Fig:PCA}, we see that high values for PC 1 is more common for communities in the politics scenarios. Conversely, most communities of the general scenario have negative $x$ coordinates, thus pointing to the opposite behavior.

A less clear picture emerges for PC 2. Large values for  PC 2 are associated with high number of replies, mentions and comments per commenter (see Figure~\ref{Fig:Explain}, right plot). For the politics scenario in Figure \ref{Fig:PCA}, communities are concentrated in the $y \in [-2,3]$ range, with those for Italy being slightly more positive than those from Brazil. In the general scenario, however, points are spread out along the $y$ axis. 

We conclude that commenters of politics influencers exhibit a more homogeneous behavior than commenters of other influencers. Particularly, commenters on politics leave larger comments and use higher tone. They also often rely on typical online social mechanisms, such as replies, mentions and emojis.
%\mt{I think this section needs some work. PCA loading scores are scores, and they reflect in the community score indirectly... I would concentrate more on saying how politics and general communities differ in terms of features...}

\subsection{Sentiment analysis}
\label{sec:sentiment}

Although communities grow around particular posts and influencers, their members do comment on posts from other influencers. Here, we analyze whether there is a difference between the sentiment expressed in comments across influencers. As explained in Section~\ref{sec:rq2}, we use SentiStrength to extract the sentiment of each comment. SentiStrength provides an integer score ranging from -4 (strongly negative) to +4 (strongly positive). Score 0 implies a neutral sentiment. We here consider as \emph{negative}, \emph{neutral} and \emph{positive} comments with scores smaller than 0, equal to 0, and greater than 0, respectively. We notice that many comments contain misspelled words and informal writing that we want to remove to prevent bias in our analysis. To this end, we use Brazilian and Italian dictionaries in Hunspell format and match the words found in the comment against them. After filtering the non-existent words, we also filter out 11\% (4\%) of the total comments for Brazil (Italy) that we discard for this analysis. 

Table~\ref{tab:fraction-comments} shows fraction of positive, neutral and negative comments. We notice that positive comments are generally more common (between 49\% and 65\%), followed by neutral comments (between 25\% and 33\%). %We look into the neutral comments to understand why they represent a significant fraction and observe a large number of short comments, misspelled words, abbreviations etc, which seem to complicate sentiment extraction by SentiStrength.
Negative comments are the minority in our data, but they are more prevalent in the politics scenarios for both countries.

\begin{table}[!b]
\footnotesize
\centering
\caption{Fraction of sentiment captured in comments using SentiStrenght.}
\begin{tabular}{cccc}
\toprule
\multicolumn{1}{|c|}{\multirow{2}{*}{\textbf{Scenario}}} & \multicolumn{3}{c|}{\textbf{Sentiment}}                                                                                 \\ \cmidrule{2-4} 
\multicolumn{1}{|c|}{}                                   & \multicolumn{1}{c|}{\textbf{Negative}} & \multicolumn{1}{c|}{\textbf{Neutral}} & \multicolumn{1}{c|}{\textbf{Positive}} \\ \midrule
\multicolumn{1}{|c|}{BR Politics}                        & \multicolumn{1}{c|}{0.12}              & \multicolumn{1}{c|}{0.26}             & \multicolumn{1}{c|}{0.62}              \\ 
\multicolumn{1}{|c|}{IT Politics}                        & \multicolumn{1}{c|}{0.18}              & \multicolumn{1}{c|}{0.33}             & \multicolumn{1}{c|}{0.49}              \\                                                                                                                                                           \hline
\multicolumn{1}{|c|}{BR General}                         & \multicolumn{1}{c|}{0.07}              & \multicolumn{1}{c|}{0.32}             & \multicolumn{1}{c|}{0.61}              \\ 
\multicolumn{1}{|c|}{IT General}                         & \multicolumn{1}{c|}{0.10}              & \multicolumn{1}{c|}{0.25}             & \multicolumn{1}{c|}{0.65}              \\ \bottomrule
\end{tabular}
\label{tab:fraction-comments}
\end{table}

% \begin{table}[!b]
% \footnotesize
% \centering
% \caption{Fraction of sentiment captured in comments using SentiStrenght.}
% \begin{tabular}{cccc}
% \toprule
% \multicolumn{1}{|c|}{\multirow{2}{*}{\textbf{Scenario}}} & \multicolumn{3}{c|}{\textbf{Sentiment}}                                                                                 \\ \cmidrule{2-4} 
% \multicolumn{1}{|c|}{}                                   & \multicolumn{1}{c|}{\textbf{Negative}} & \multicolumn{1}{c|}{\textbf{Neutral}} & \multicolumn{1}{c|}{\textbf{Positive}} \\ \midrule
% \multicolumn{1}{|c|}{BR Politics}                        & \multicolumn{1}{c|}{0.10}              & \multicolumn{1}{c|}{0.32}             & \multicolumn{1}{c|}{0.58}              \\ 
% \multicolumn{1}{|c|}{IT Politics}                        & \multicolumn{1}{c|}{0.13}              & \multicolumn{1}{c|}{0.40}             & \multicolumn{1}{c|}{0.47}              \\                                                                                                                                                           \hline
% \multicolumn{1}{|c|}{BR General}                         & \multicolumn{1}{c|}{0.06}              & \multicolumn{1}{c|}{0.41}             & \multicolumn{1}{c|}{0.53}              \\ 
% \multicolumn{1}{|c|}{IT General}                         & \multicolumn{1}{c|}{0.04}              & \multicolumn{1}{c|}{0.36}             & \multicolumn{1}{c|}{0.57}              \\ \bottomrule
% \end{tabular}
% \label{tab:fraction-comments}
% \end{table}

\begin{figure}[!t]
    \begin{subfigure}{0.9\columnwidth}
        \centering
        \includegraphics[width=0.9\columnwidth]{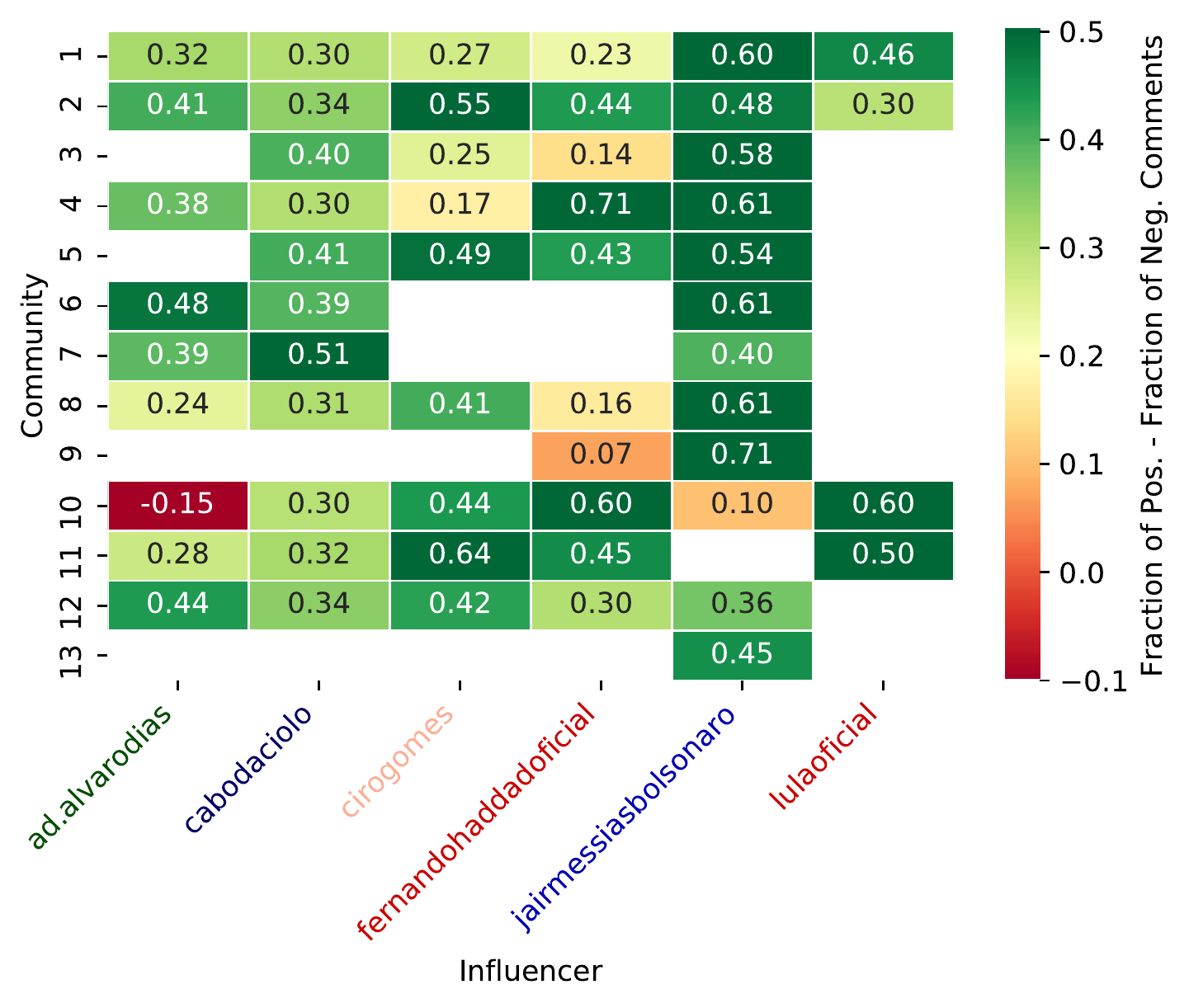}
        \caption{Brazil (week 5)}
        \label{Fig:sentiment_BR}
    \end{subfigure}
    \begin{subfigure}{0.9\columnwidth}
        \centering
        \includegraphics[width=0.9\columnwidth]{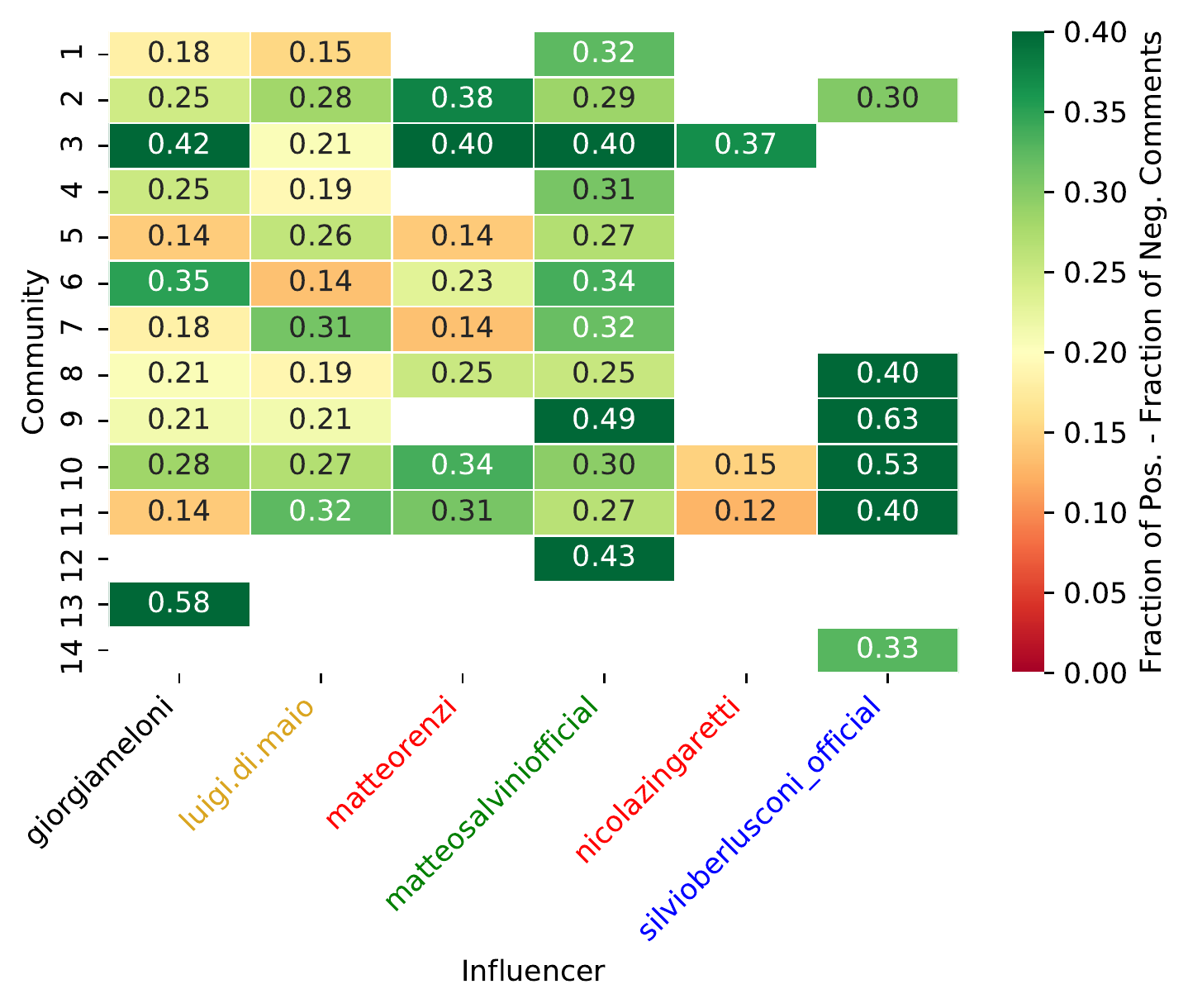}
        \caption{Italy (week 7)}
        \label{Fig:sentiment_IT}        
    \end{subfigure}
    \caption{Contrastive sentiment score (difference between fraction of positive and negative comments) of communities towards political leaders during the main election week.}
    \label{Fig:sentiment} 
\end{figure}

We now analyze how the communities' sentiment varies towards profiles of different politicians. More specifically, we compute the breakdown of positive, neutral and negative comments of each community on posts of each influencer. To summarize differences, we report in Figure~\ref{Fig:sentiment_BR} and Figure~\ref{Fig:sentiment_IT} a \emph{contrastive score} calculated as the difference between the fractions of positive and negative comments for the particular community and influencer. We ignore cases where a community has made less than 100 comments on a given influencer's posts to ensure that samples are representative. These cases are marked as white cells in the heatmaps.\whiteColor{\ref{tab:political-words}}

\begin{table*}[!h]
\centering
\scriptsize
\caption{Example of words with the highest TF-IDF for some communities in the politics scenario in the main election week.}
\label{tab:political-words}
\begin{tabular}{|c|c|c|c|}
\toprule
\textbf{Scenario} & \textbf{Comm.} & \textbf{Key Words}                                                                                                                              & \textbf{Context}                                                                                                                                                           \\ \midrule
BR  & 3                  & \begin{tabular}[c]{@{}c@{}}`Anapolis', `Orla', `Righteousness', `Constant', \\ `Natal',  `Paulista', `Spontaneous', \\ `JB17', `Gesture', `Avenue' \end{tabular} & \begin{tabular}[c]{@{}c@{}}It refers to several places \\where pro-Bolsonaro rallies  took \\ place during the election campaign.\end{tabular}                                    \\ \hline
BR  & 7                  & \begin{tabular}[c]{@{}c@{}}`Nazi', `Jew', `Hitler', \\ `Black People', `Anonymity', `Bozonazi', \\`Distrust', `Jerusalem', `Homosexual' \end{tabular}          & \begin{tabular}[c]{@{}c@{}}It refers to Bolsonaro's posts about \\ specific social groups in an attempt to show \\ he has no prejudice against such groups.\end{tabular} \\ \hline

BR  & 10                  & \begin{tabular}[c]{@{}c@{}}`Manuela', `Haddad, `Scammer', \\ `Lulalivre', `Birthday', `Guilherme', \\ `Dilma', `Gratefulness', `Lula'\end{tabular}          & \begin{tabular}[c]{@{}c@{}}It refers to left-wing names, \\ such as Fernando Haddad,  his \\ deputy Manuela, Dilma Rousseff and Lula (ex-presidents).\end{tabular} \\ \hline

BR  & 11                 &
\begin{tabular}[c]{@{}c@{}}`Ciro', `Experience', `Political Activism', \\ `Polarization',  `Brazil', `Second Round', \\`Turn', `Prepared', `Project' \end{tabular}    & \begin{tabular}[c]{@{}c@{}}It refers to the center-left candidate \\ Ciro Gomes  who arrived close to \\reach the second round of the elections.\end{tabular}                                                                                          \\ \midrule %\hline \hline
IT  & 3                  & \begin{tabular}[c]{@{}c@{}}`Gooders', `Big ciao'', `Captain', \\`Crime', `Good night', `Polls', \\ `Never Give Up', `Electorate', `Lampedusa', `Riace' \end{tabular} & \begin{tabular}[c]{@{}c@{}}General Salvini's jargon, \\as well as places  related to \\the arrival of immigrants in Europe (e.g., Lampedusa).  \end{tabular}                                                                                                        \\ \hline
IT  & 4                  & \begin{tabular}[c]{@{}c@{}} `Monetary', `Elite', `Unity', `Budget',\\ `Fiscal', `Colonial', `Equalize', \\`Yellow Vests', `Masonic', `Store', `IVA', \end{tabular} & \begin{tabular}[c]{@{}c@{}}Generic taxes and monetary issues.\end{tabular}                                                                                                        \\ \hline
IT  & 10                  & \begin{tabular}[c]{@{}c@{}}`Consumption', `Fuel', `Insurance', `Traffic',\\ `Helpless', `Vehicular', `Taxes', `Redundancy', \\`Veterinary', `Animal rights', `Cats', `Abuse', `Cruelty', `Breeding'\end{tabular} & \begin{tabular}[c]{@{}c@{}} A combination of terms related\\ to taxes, vehicles and animals' rights.\end{tabular}                                                                                                        \\ \hline
IT  & 11                  & \begin{tabular}[c]{@{}c@{}}`5S', `Toninelli (ex-Transport Minister)', `Corruption',\\ `Zingaretti (PD's leader)', `Calenda (ex-PD politician)', \\`Honesty', `Election list', `Coalition', `Budget', `Growth'  \end{tabular} & \begin{tabular}[c]{@{}c@{}}Debate on Five Stars Movement (a government\\ party at the time) and Democratic \\Party (the main opposition party at the time)\end{tabular}                                    \\ \bottomrule
\end{tabular}
\end{table*}

In Figure~\ref{Fig:sentiment_BR} we consider the six political leaders already used for Figure~\ref{fig:hetmaps}. We focus on the week of the first election round in Brazil (week 5). Predominantly, communities make positive comments on the profiles in which they are more active, i.e., their ``referring candidate''. More negative comments are seen on ``opposing candidates''. For instance, communities 1 to 9, highly active on Jair Bolsonaro's posts, display a more positive contrastive score on them. Analogously, communities 10 to 12, mostly formed by commenters very active on the profiles of left-wing influencers such as Ciro Gomes (cirogomes) and Fernando Haddad (fernandohaddadoficial), tend to write negative comments on their opponents, such as Jair Bolsonaro. This behavior appears on all weeks and suggests that communities in politics tend to promote their central influencers while trying to demote others.

Considering the Italian case, we observe similar results in Figure~\ref{Fig:sentiment_IT}. Communities exhibit positive contrastive scores towards candidates in general, but with higher scores for the referring candidate.

\subsection{Main topics of discussion}
\label{sec:res_textual}

We now turn our attention to the analysis of the main topics around discussions. As before, we focus on politics, during the election weeks. To summarize the overall behavior of each community, we group together all their respective comments in one document. As explained in Section \ref{sec:rq2}, the documents build a corpus on which we then use the TF-IDF metric to identify the most representative words of each document (i.e., community), henceforth called \emph{top words}.

We show in Table~\ref{tab:political-words} the top-10 words (translated to English) for communities yielding the most interesting observations. We manually inspect the comments and related posts, providing a reference context as the last column of the table. The manual inspection suggests that these words give a good overview of what the communities discuss. 

Matching with Figure~\ref{fig:examples}, communities 3 and 7 for Brazil are associated with rallies in different locations in the country, and with debunking Bolsonaro's prejudice against ethnic and racial groups. The terms highlighted by TF-IDF reflect quite accurately the respective topics, reporting locations of rallies and words linked to racism and Nazism. Similarly, the top words for communities 10 and 11 are focused on the names of the candidates whose profiles they mostly comment on. For Italy, community 3 reflects the typical jargon used by Salvini's supporters. Community 4 debates on taxes and monetary issues. Community 10's comments refer to provoking posts that mix taxes, car costs and animals' rights. Last, community 11 seems to debate over the left-wing party (the main opposition party at the time) and the 5-Stars movement (the governing party at the time). 

In a nutshell, the TF-IDF is instrumental to analyze what the communities are discussing. The analysis demonstrates that communities are well-formed around the topics they discuss, even if they have been built solely on the network of commenters' interactions.

\subsection{Psycholinguist properties}

In this section, we study the psycholinguistic properties of comments, aiming at finding similarities and differences in the way commenters of communities communicate. We rely on the Linguistic Inquiry and Word Count (LIWC) tool to calculate the degree at which various categories of words (called attributes in the LIWC terminology) are used in a text (see Section~\ref{sec:rq2}). For example, attribute \emph{Home} includes the words ``Kitchen'' and ``Landlord, and attribute \emph{Family} ``Daughter'', ``Dad'' and ``Aunt''. 

For each community, we run LIWC on the comments and compute the average frequency of the attributes. We then look for statistical differences between communities based on the average values of the attributes. For Brazil, we identify 62 attributes (from the 64 available in LIWC's Portuguese dictionary) for which differences across communities are statistically significant\footnote{We used the Kruskal non-parametric test to select attributes with a significant difference between the distribution of community comments considering $ p-value = 0.01 $.}. For Italy, we identify 77 (from 83 available in the LIWC Italian dictionary). From those, we select the five attributes that exhibit the largest variability across communities in terms of Gini index and use them characterize the psycholinguistic of communities. 

\begin{figure}[!t]
    \centering
    \subfloat[Brazil, Politics.]{\includegraphics[width=.88\columnwidth]{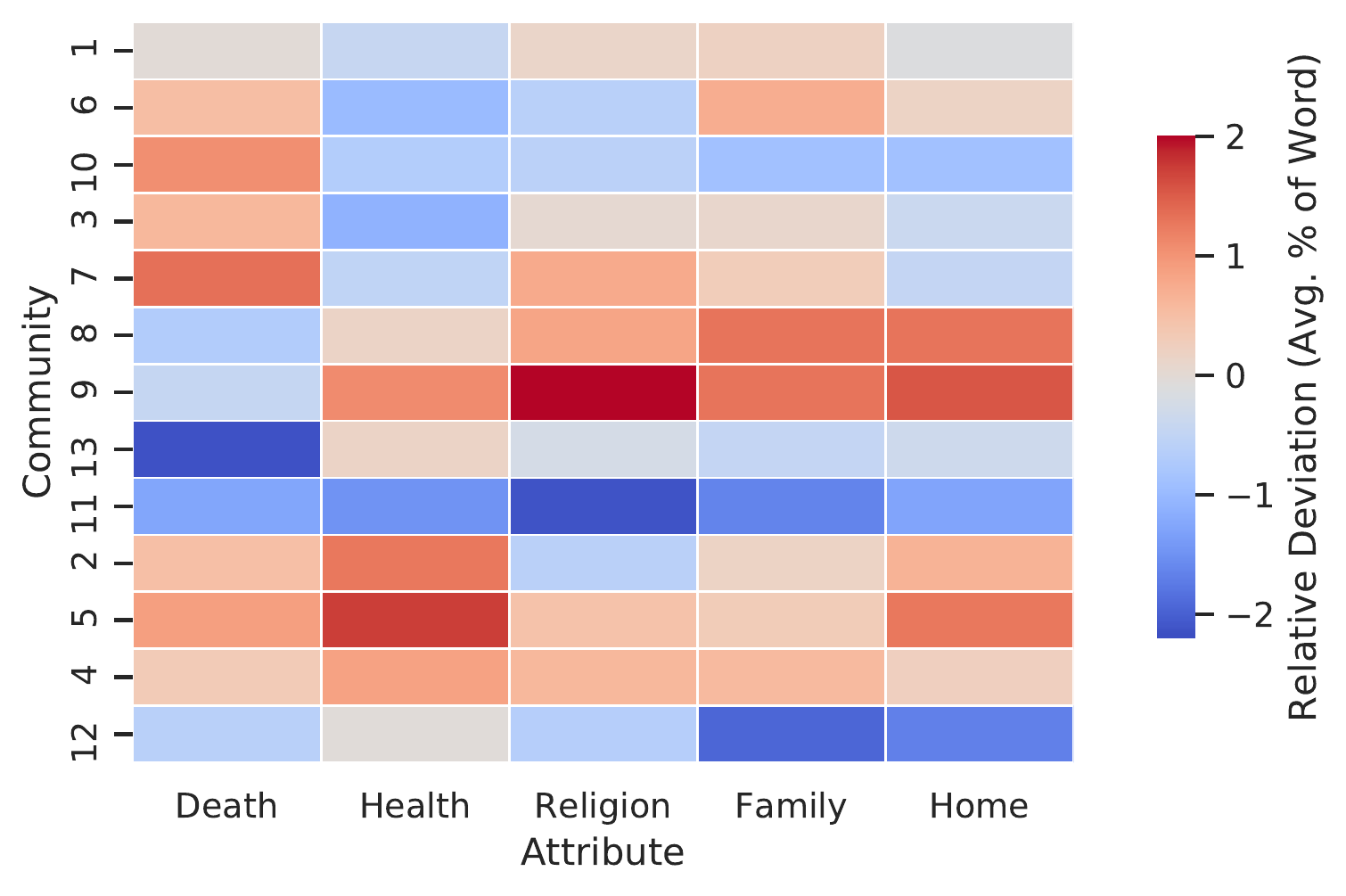}\label{Fig:Liwc_BR_politics}}
    \qquad
    \subfloat[Italy, Politics.]{\includegraphics[width=.88\columnwidth]{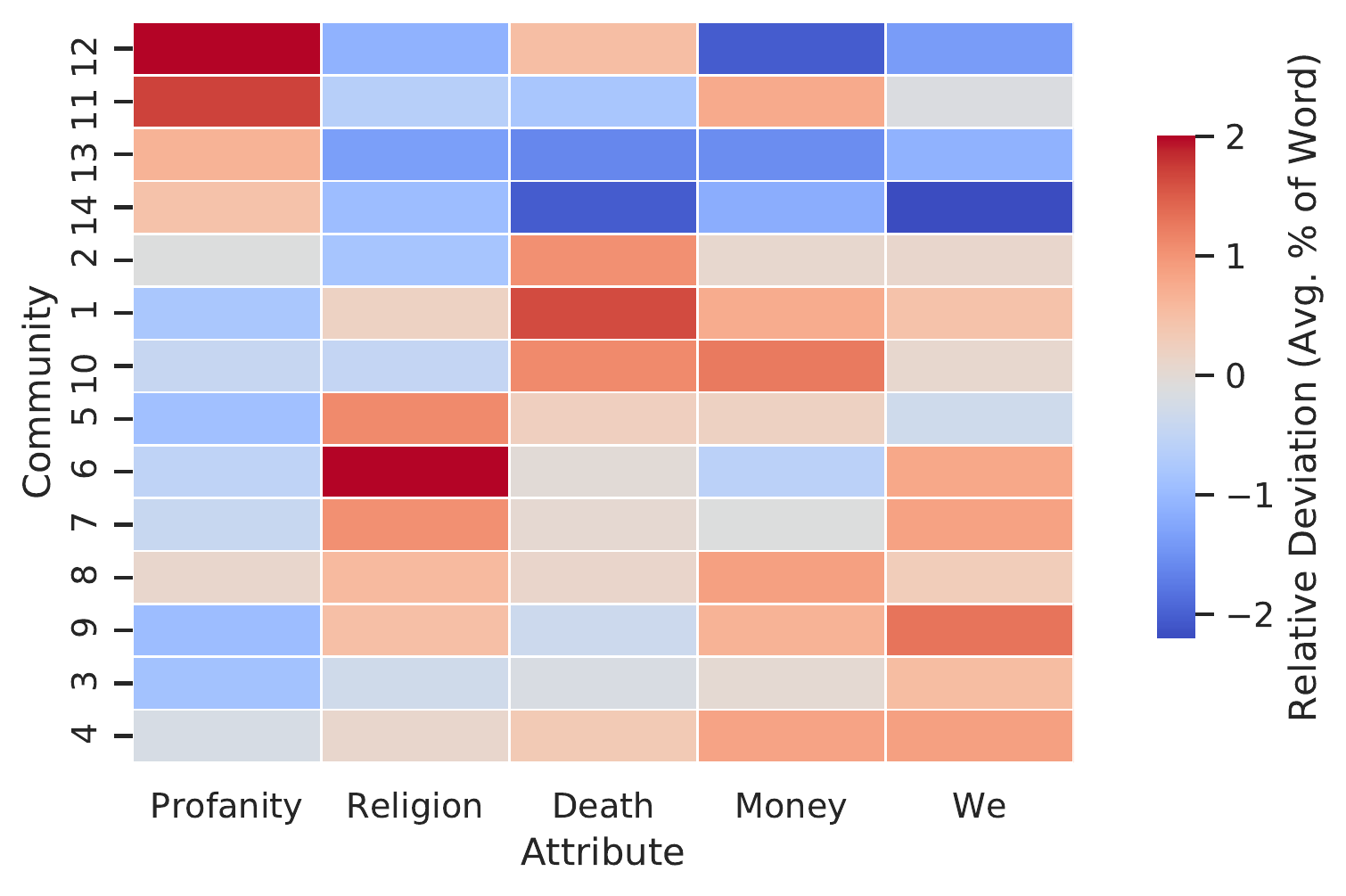}\label{Fig:Liwc_IT_politics}} \\ \caption{ Top 5 LIWC attributes and their relative difference between communities.}\label{Fig:Liwc_politics}
\end{figure}

Figure~\ref{Fig:Liwc_politics} shows heatmaps for the top-five attributes found for the Brazilian (top) and Italian (bottom) politics scenarios. The heatmap cells in a column indicate the relative deviation of the given attribute for the given community from the other communities. In other words, each column (attribute) is z-score normalized -- i.e., $z = (x - mean)/std$. Thus, each value gets subtracted the average of the column, then divided by the standard deviation of the column. The results show how communities are different in terms of the LIWC selected attributes. For instance, for Brazil, Politics, communities 6, 10, 3 and 7 frequently use words regarding \emph{death}, but seldom words related to health. Communities 2, 5 and 4 show positive scores on almost all attributes. Community 13 focuses mostly on \emph{health}. In Italy, community 6 is very focused on \emph{religion} (commenters debated Salvini's post that depicts a Rosary). Communities 3, 4, 6, 7, 8 and 9 adopt attributes \emph{we} representing words and verbs in the first person plural (e.g., we, our and us). This kind of word is used used in phrases aiming at aggregating the community towards the same goal. Community 12 and 13 exhibit some hate speech \emph{}.

In summary, LIWC is a useful tool to analyze the content of Instagram comments, complementing the TF-IDF analysis with information on the topics being debated. We find that communities debate on different topics and using different lexicon. \whiteColor{\ref{fig:Activity}} 

\section{Temporal analysis}
\label{sec:rq3}

In this section, we focus on the dynamics of communities during the 10 weeks of observation. First, we analyze the community membership, studying to what extent commenters persist in the network backbone and are found in the same communities across weeks. Next, we characterize the dynamics of the content, i.e., the topics that these communities are engaged in.

\subsection{Community membership persistence}

We start our analysis by studying the persistence of commenters inside the  network backbone and to what extent these commenters end up in the same community week by week. We also want to check if the most engaged commenters exhibit a different behavior -- i.e., tend to persist more than those who are less engaged. To this end, we perform a separate analysis selecting the top-1\% and top-5\% commenters in terms of number of comments in week $w$ and $w+1$. Then, we compute the persistence and NMI score (see Section~\ref{sec:metho_tempo}), restricting to these commenters and comparing the results with those obtained with the full set of commenters.

\begin{figure}[!t]
    \centering
    \subfloat[Brazil - Politics.]{\includegraphics[width=0.49\columnwidth]{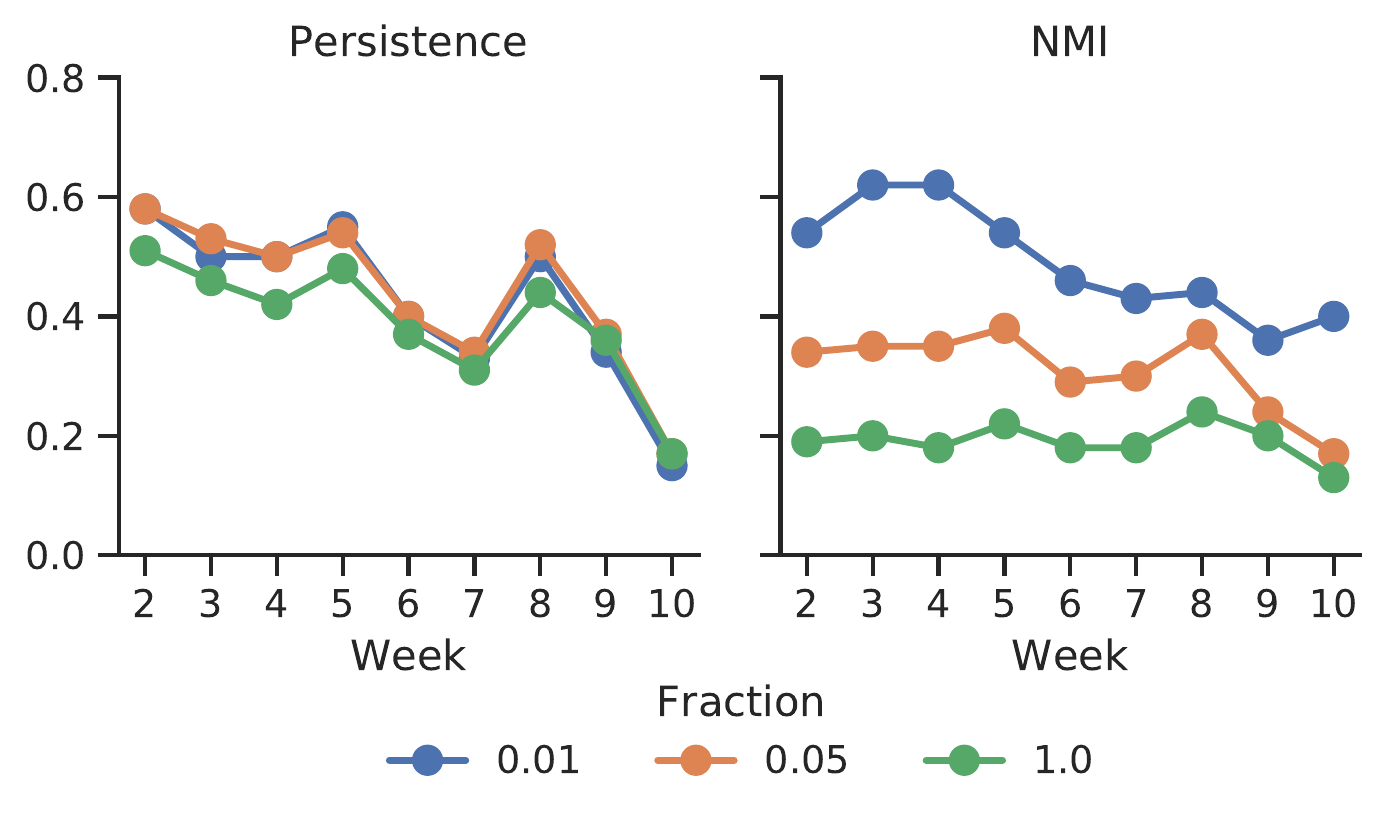}\label{Fig:Temp_BR_politics}\hspace{0.1cm}}
    \subfloat[Italy - Politics.]{\includegraphics[width= 0.49\columnwidth]{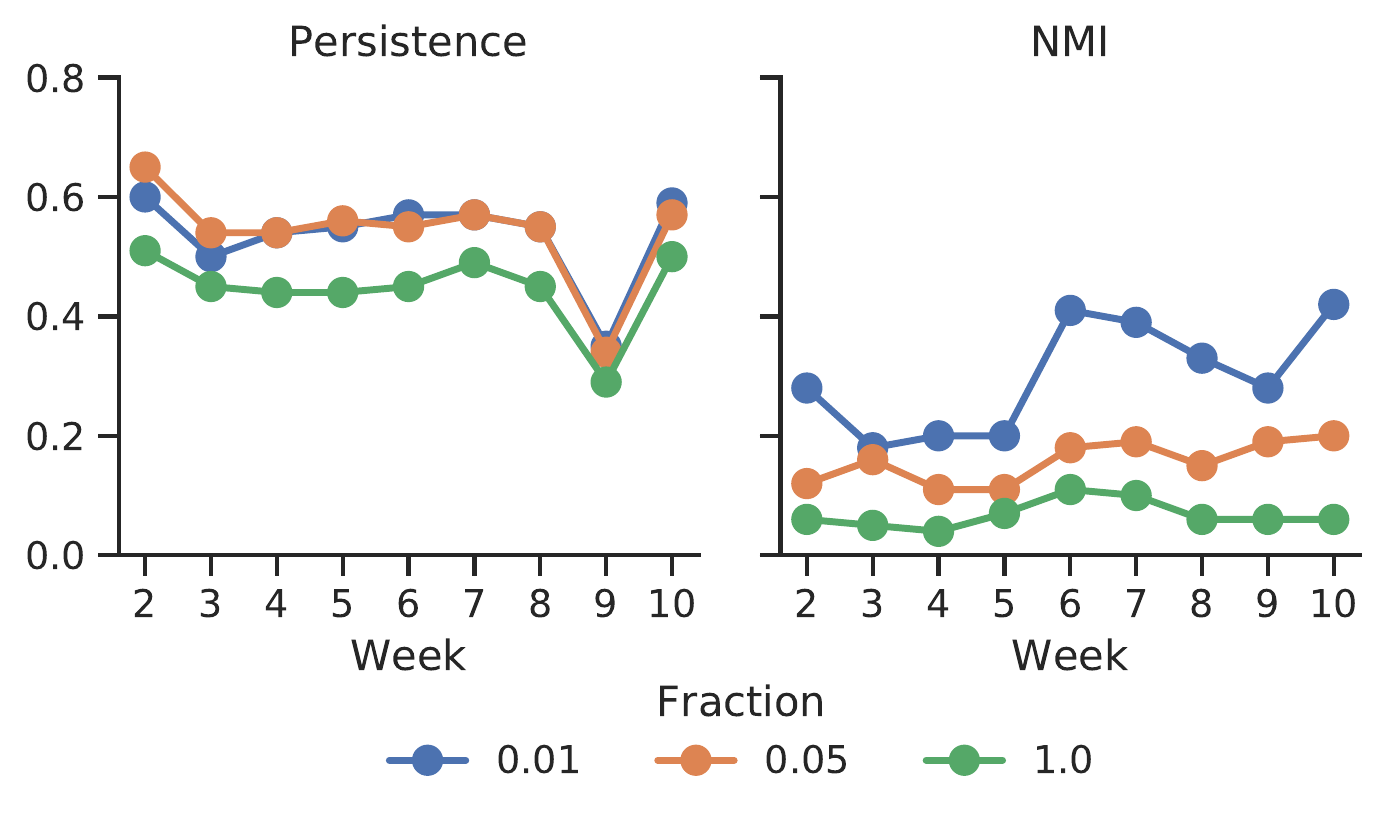}\label{Fig:Temp_IT_politics}} \\ \vspace{0.5cm}
    \subfloat[Brazil - General.]{\includegraphics[width= 0.49\columnwidth]{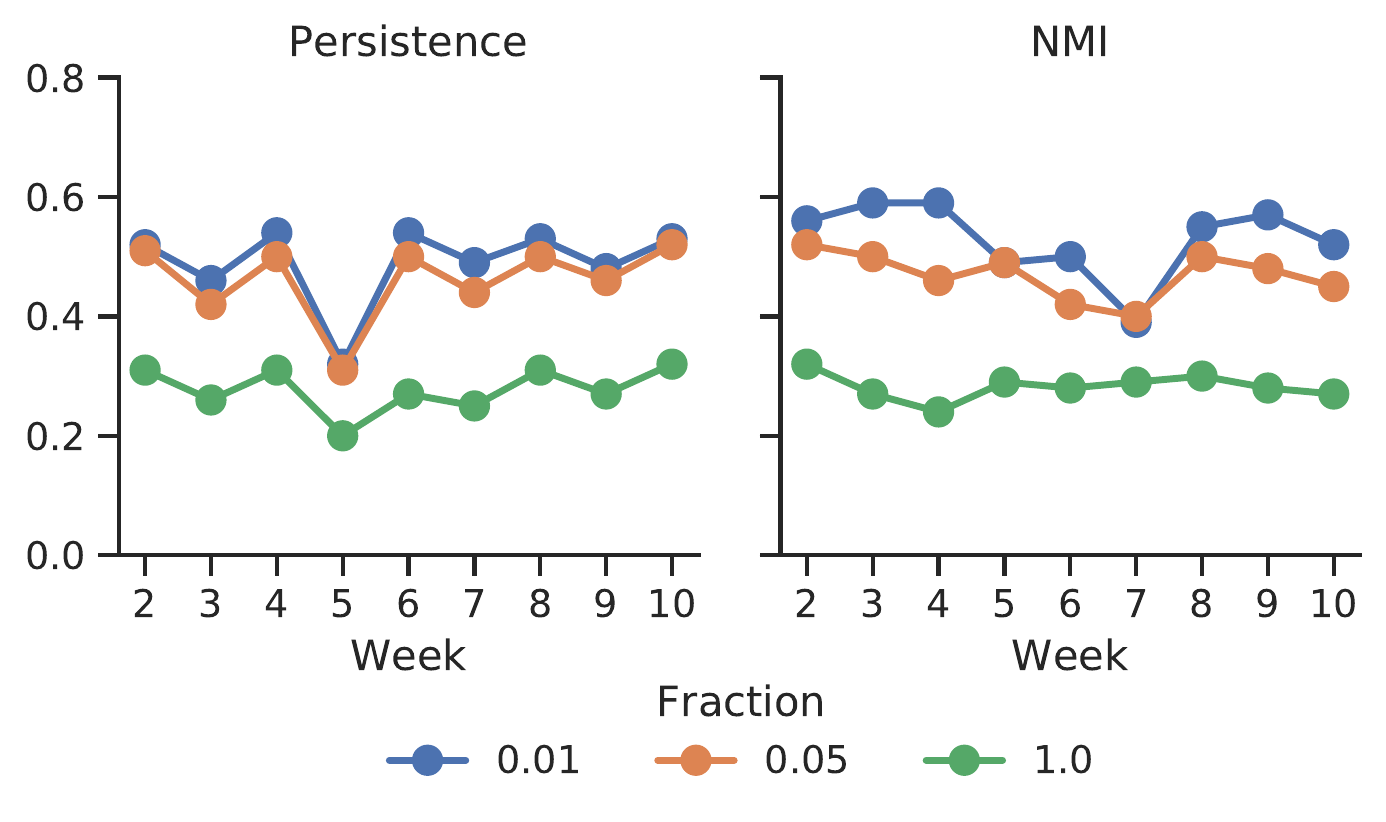}\label{Fig:Temp_BR_general}\hspace{0.1cm}}
    \subfloat[Italy - General.]{\includegraphics[width= 0.49\columnwidth]{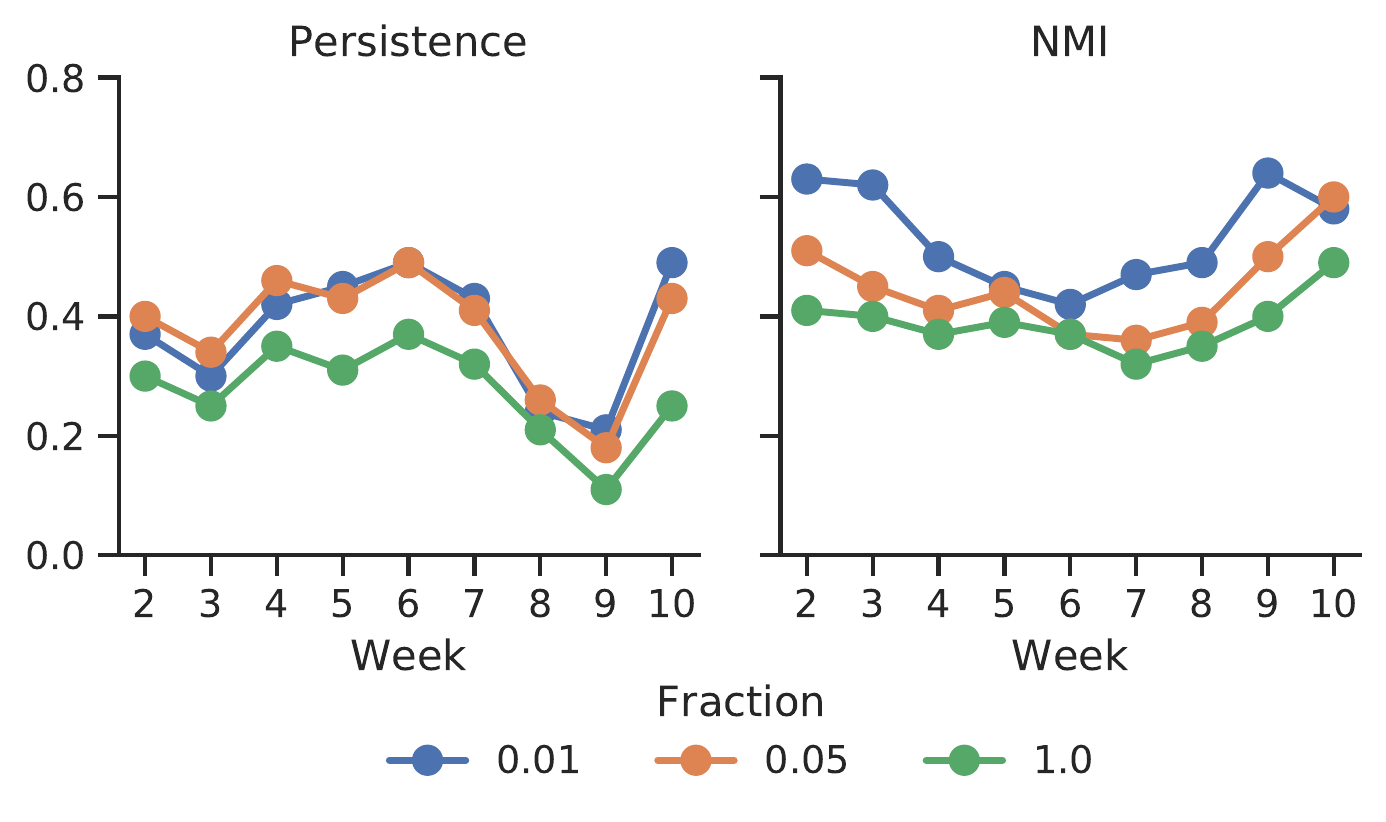}\label{Fig:Temp_IT_general}}
    \caption{Temporal evolution of commenters in communities. Blue: top 1\%, Orange: top 5\%, Green: all commenters.}\label{Fig:General_NMI_Pers}
\end{figure}

We report results in Figure~\ref{Fig:General_NMI_Pers} separately by country and for Politics and General. Considering Politics (Figures~\ref{Fig:General_NMI_Pers}a and \ref{Fig:General_NMI_Pers}b), we note that the persistence in Brazil is moderately high, regardless the subset of commenters. Around 50-60\% of commenters remain in the backbone week after week until the first round of elections (week 5). Since then, we observe a decrease (also due to the drop of commenters in general) until the second round election (week 8), followed by a significant drop after. This trend shows that commenters were very engaged in the election period, mostly in the first round when the debate included more politicians, senators, congressmen and governors. In the second round, fewer candidates faced -- yet people were consistently engaged before finally plumbing two weeks after elections. These results corroborate the first intuition we observed in Table~\ref{tab:charact} -- where the number of commenters varied over time. Since persistence is similar for all subsets of commenters, we can conclude that all commenters in the backbone are persistently engaged. That is, the backbone members are quite stable.

Considering the membership of commenters within the same community, the NMI shows that the top-1\% and top-5\% most active commenters (blue and orange curves) are considerably more stable in their communities during the whole time. When considering all commenters in the backbone, the NMI is significantly lower. This is due to the birth and death of new communities, centered around specific topics, where the debate heats up and cools down. These dynamics attract new commenters that afterward disappear or change community.

For Italy, Politics (Figure~\ref{Fig:General_NMI_Pers}b) different considerations hold. The constant persistence suggests a stable engagement of commenters in the backbone. We just observe a sudden drop the week after the election, where the interest in the online debate vanished. On the other hand, the NMI is rather low, revealing more variability in community membership, even if we restrict our attention to the most active commenters. Despite commenters in the backbone tending to be the same (persistence is typically above 0.5), they mix among different communities. Considering the low modularity of communities for this scenario (see Table \ref{tab:sum_backbone_networks}), we conclude that the community structure is weaker in this case, indicating overlapping among communities that favor membership changes. This result is also visible from the dendrogram in Figure~\ref{fig:dendo}, where we observe that influencers receive comments from similar communities making the latter also more clustered.

Moving to General (Figures~\ref{Fig:General_NMI_Pers}c and \ref{Fig:General_NMI_Pers}d), we observe slightly lower persistence than in Politics, but more stable over time. NMI instead often results higher for General than Politics, reflecting better separation between communities, which persist over time. More in detail, for Brazil (Figure \ref{Fig:General_NMI_Pers}c) we observe that persistence and NMI are high and stable -- especially for the most active users. This suggests that the most engaged commenters have diverse, specific and stable interests. Indeed, here there is no exogenous event that pushes a temporal dynamic, like elections do for politics. Again, this result reflects the high heterogeneity of posts and influencers in the General category. Moving to Italy, Figure \ref{Fig:General_NMI_Pers}d shows that persistence is small and varies over time. Here the lower popularity of Instagram in Italy than in Brazil may play a role, coupled with the smaller number of comments (see Table~\ref{tab:charact}). However, NMI is high and stable. We conclude that although many users do not persist in the backbone, the remaining are very loyal to their interests.

\subsection{Topic persistence}

\begin{figure}[t]
    \centering
    \subfloat[Brazil - Politics.]{\includegraphics[width=0.8\columnwidth]{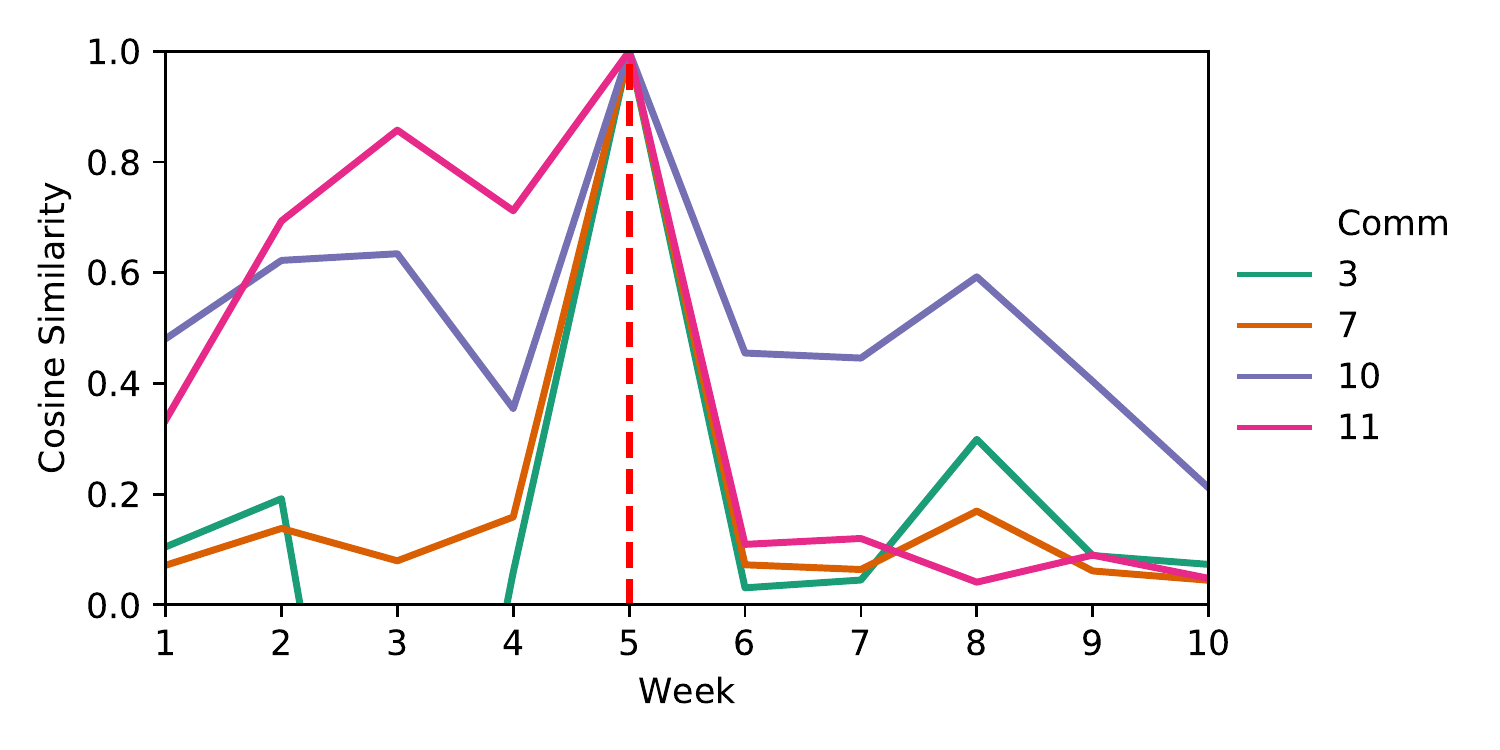}\label{Fig:TF-IDF_Temp_BR}\hspace{0.1cm}}\\
    \subfloat[Italy - Politics.]{\includegraphics[width= 0.8\columnwidth]{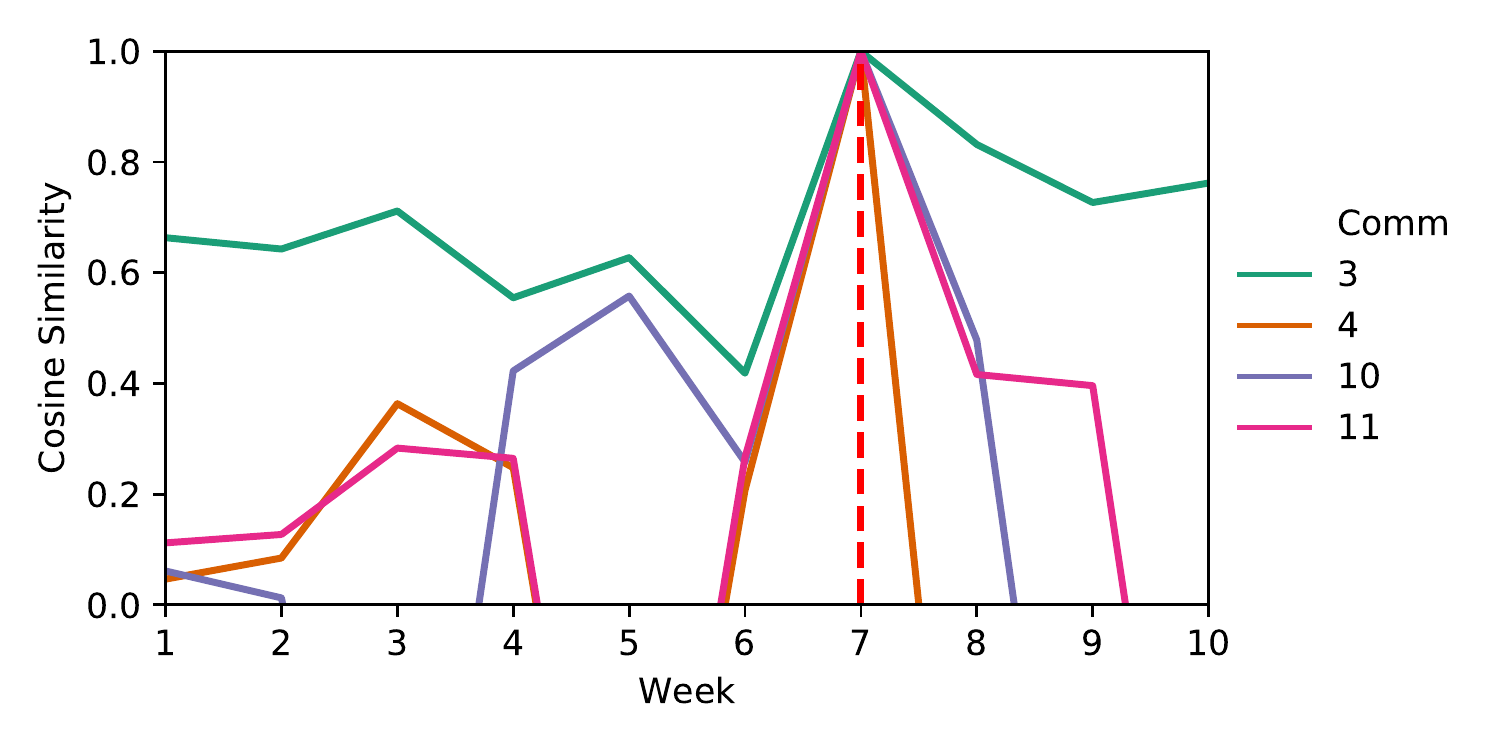}\label{Fig:TF-IDF_Temp_IT}}
    \caption{Example of how communities' comments change over time. We set weeks 5 and 7 as reference, being the election weeks in Brazil and Italy, respectively. }\label{Fig:TF-IDF_temporal}
\end{figure}

We now discuss how the topics discussed by communities evolve over time. To that end, we take as reference  weeks 5 for Brazil and 7 for Italy in the political scenario, being the weeks of elections in each country. We compute the cosine similarity between the communities in the reference weeks (illustrated in Table \ref{tab:political-words}) and the communities extracted in all other weeks, for each country. 
%$w \in [1, 10]$ summarized in Table \ref{tab:political-words}.   \ju{Nao entendi referencia a Tabela 6 aqui. Ela nao tem weeks.  w e week, nao? Ou entao a notacao ta errada. Acho que e esta frase ai ta bem misleading.} 
That is, for a given week, we identify whether there exists a document/community that is significantly similar to those found in the reference week, following the steps presented in Section \ref{sec:metho_tempo}. 

Figures~\ref{Fig:TF-IDF_temporal} show examples for both scenarios. In weeks 5 and 7 for Brazil and Italy, respectively, the cosine similarity is 1 since the documents are compared with themselves. Focusing on Brazil first, we observe a very distinct behaviors among the picked-up examples. Remember that communities 3 and 7 are focused, mainly, on Bolsonaro's profile and comment on posts related to \textit{rallies} and \textit{racism}, respectively. In both cases, we can observe that discriminating terms for these communities are momentary and sporadic, with some communities using terms about \textit{rallies} that appear in some weeks, still with a very similarity low. Conversely, the set of significant terms representing community 10 and related to candidate Fernando Haddad. At last, consider community 11, focused on Ciro Gomes. Again, we can observe that terms used by his community in the election week were used in some communities earlier, exhibiting high similarity. However, immediately after the first round (week 5) when Ciro Gomes lost, the similarity drops significantly. Indeed, Ciro Gomes was excluded from the run-off and the online debate (and community) suddenly vanished.
In Italy (Figure \ref{Fig:TF-IDF_temporal}b) we observe a similar behavior. Community 3, mostly consisting of Salvini’s supporters, use a very specific jargon and are always present. Community 4 debates around taxes and monetary issues, which already debated during week 3. The same considerations hold for community 10, in which the fight between democrats and five stars supporters heats more frequently.

In summary, people commenting in politics are more volatile than those commenting on general topics, with debates suddenly growing and cooling down. Some sets of terms remain ``alive'' throughout the observation period, while other include communities born around short events such as rallies, which take place on a specific date.

\section{Discussion on main findings and conclusions}
\label{sec:discussion}

Our work contributes with a deep analysis of interactions on Instagram. Here, we discuss the major findings of our analysis and put them in perspective of the relevant literature in the field. Among key insights, we can mention:

\begin{itemize}
    \item We introduced a reference probabilistic network model to select salient interactions of co-commenters on Instagram. We found that most users have at least one interaction that deviates from the reference model, similarly to previous work on user behavior on online social networks~\cite{Burke:2009, Wilson:2009, Kwon:2014}. Our model allows us to extract better quality information and thus better understand how the political debate evolves on Instagram.
    
    %However, we found higher percentages of salient interactions in politics -- 84\% (87\%) for Brazil (Italy) -- than in other categories -- 65\% (60\%) for Brazil (Italy). These results suggest that influencers in politics can mobilize more commenters, even if they usually post less and are followed by smaller number people than other influencers.
    
    \item Commenters exhibiting salient interactions build strong communities, often co-interacting on posts from different influencers. Hambrick et al.~\cite{Hambrick:2014} analyzed tweets about the 2012 London Summer Games, finding that communities are born around viral posts. In politics, we reveal that multiple communities emerge around different subsets of posts of the same influencer, and this behavior seems to be guided by the discussed topics on Instagram as well. 
    
    \item Comments produced by political communities tend to be longer, richer in emojis, hashtags and uppercase words, indicating assertive and emotional content than communities around other topics. Kim et al.\cite{Kim:2020} analyzed multiple categories of Influencers (outside politics) finding that such textual features are not very discriminative across the categories. We confirm their findings on our baseline of non-politicians, however finding remarkable differences for Influencers on politics.

    \item We compared the use of hashtags and mentions across categories of influencers. Previous works~\cite{Trevisan:2019, Yang:2019, Kang:2020} associated these aspects to information spreading strategies. We concluded that such strategies are even more prominent on politics. 
    
    \item We show a predominance of positive sentiment, confirming findings of \cite{Zhan:2018, Arslan:2019}. We observe also larger percentages of negative comments on politics, a phenomenon also studied by Kuvsen et al.\cite{kuvsen:2018} for dissemination of information during the 2016 Austrian elections. However, here we also find that communities built around a specific politician tend to leave negative comments on profiles associates to opposite political spectra. 
    
    \item In weeks preceding elections, we noticed a heating up in political debate and large variations on community membership, which is reduced after the election day. Recurrent behaviors on Instagram had so far been evaluated only in other context, such as cyberbullying~\cite{Cheng:2020, Gupta:2020, Cheng:2021}. 
    
    \item We observe a large variety diversity in discussed topics over time for communities in politics. Whereas some topics attract attention only momentarily (e.g., racism), others, centered around more fundamental fundamental political subjects (e.g., rallies, particular candidates and political ideologies), remain consistently active over time. The rapid decline of communities around specific topics has been observed on Twitter as well~\cite{moody2019analysis}.
\end{itemize}

Our analyses open up a number of potential avenues for further exploration. One possibility is extending the study to other platforms, such as Facebook and Twitter, with a similar network structure, which could reveal distinguishing patterns. In particular, it would be interesting to compare the dynamics of discussions around posts by the same person on different platforms, to shed lights on the role of different social media platforms in current society. 

Another interesting future work is the evaluation of the role of  commenters on each community -- e.g., considering topological and activity-related attributes to determine commenters' importance. Similarly, another interesting direction is the investigation of the role of vertices in the backbone (and in particular communities) as part of information dissemination campaigns, notably misinformation campaigns.

% \ch{Application of this model in other OSNs with the same structure such as Facebook and Twitter to observe how the network structure as well as the characteristics analyzed change. Several of these applications have an identical or similar structure, which makes the application of our model feasible.  ----- \textbf{(I dont know if it makes sense. Just ideas.)}  Moreover, is it possible to identify parallel behaviour in both or more of these platforms on specifics profiles? For instance, mainly in politics, the influencers post the same media in two or more of these platforms. How do users behave on distinct platforms? Hos does debate essence change on them? Also, a similar analysis can be done for Whatsapp based on a paper that I 'read'. The authors did something to understand how posts that had its links shared in groups of WhatsApp received comments after the sharing. }

% \ch{ Communities in multilayers networks - Each layer one OSN - How to identify the same users in many OSNs? 

% Olhar para os nós centrais -> In politics, analysis of users ideologies, supporters and opposers to profiles.

% Miss information spreading - How does users in a backbone contribute to miss information spreading? }

\section*{Acknowledgement}

The research leading to these results has been funded by the SmartData@PoliTO Center for Big Data technologies at Politecnico di Torino, Brazilian National Council for Scientific (CNPQ), Technological Development and Brazilian Coordination for the Improvement of Higher Education Personnel (CAPES) and Minas Gerais State Foundation for Research Support (FAPEMIG).

\bibliographystyle{elsarticle-num}

\bibliography{mybibfile}

\end{document}